# A validated energy model of a solar dish-Stirling system considering the cleanliness of mirrors

Alessandro Buscemi[1]*, Valerio Lo Brano[2], Christian Chiaruzzi[3], Giuseppina Ciulla[2], Christina Kalogeri[4] ☼

[1] Graded S.p.A. Naples, Italy; alessandro.buscemi@graded.it
[2] Department of Engineering, University of Palermo, Italy; valerio.lobrano@unipa.it; giuseppina.ciulla@unipa.it;
[3] Horizonfirm S.r.l, Palermo, Italy; c.chiaruzzi@horizonfirm.com
[4] National & Kapodistrian University of Athens, Department of Physics, Greece, chriskal@mg.uoa.gr
* Correspondence: alessandro.buscemi@graded.it

**Abstract**

Solar systems based on the coupling of parabolic concentrating collectors and thermal engines (i.e. dish-Stirling systems) are among the most efficient generators of solar power currently available. This study focuses on the modelling of functioning data from a 32 kWe dish-Stirling solar plant installed at a facility test site on the University of Palermo campus, in Southern Italy. The proposed model, based on real monitored data, the energy balance of the collector and the partial load efficiency of the Stirling engine, can be used easily to simulate the annual energy production of such systems, making use of the solar radiation database, with the aim of encouraging a greater commercialisation of this technology. Introducing further simplifying assumptions based on our experimental data, the model can be linearised providing a new analytical expression of the parameters that characterise the widely used Stine empirical model. The model was calibrated against data corresponding to the collector with clean mirrors and used to predict the net electric production of the dish-Stirling accurately. A numerical method for assessing the daily level of mirror soiling without the use of direct reflectivity measures was also defined. The proposed methodology was used to evaluate the history of mirror soiling for the observation period, which shows a strong correlation with the recorded sequence of rains and dust depositions. The results of this study emphasise how desert dust transport events, frequent occurrences in parts of the Mediterranean, can have a dramatic impact on the electric power generation of dish-Stirling plants.

1. Introduction

The rise in carbon emissions, to a large extent due to the combustion of fossil fuels to generate electricity, has been well-documented over the past few decades, with many studies indicating the need for innovation in terms of renewable energy generation and storage [1]. Indeed, the rising global demand for electricity, which is predicted to continue, represents an enormous environmental risk, and how we choose to generate this electricity will be a key factor in determining the future energy policy of the European Union [2].

Concentrated Solar Power (CSP) systems are among the most modern renewable energy systems currently used to harness solar energy [3], with near zero carbon emissions [4]. These CSP systems operate by using a number of mirrors to concentrate the sun's rays on a relatively small receiver. The resulting heat can be used either to produce steam to power a steam turbine for electricity generation or directly to power industrial processes [5]. There are four CSP technologies already being used globally in the heat and power industry: parabolic trough concentrators, linear Fresnel reflectors, solar towers, and parabolic dish concentrators [3], [6].

Compared to other CSP systems, parabolic dish technology has demonstrated the highest levels of efficiency in the conversion of solar to thermal energy [7], [8], [9], [10] using a parabolic concentrator which tracks the sun and concentrates the direct component of the solar radiation on a receiver fixed at its focal point. The high-temperature heat in the receiver is transferred to a working fluid in the Stirling engine, which then produces mechanical energy [11], [12]. The mechanical energy is subsequently transformed into electricity by a generator. Existing plants use Stirling cycle engines powered by air or hydrogen [9], [10], while recent prototype plants have been realised with the use of air microturbines. Some advantages of the dish-Stirling system are that it has high cycle thermal efficiency, fewer moving parts than other systems of similar dimensions, and a smooth-running, quiet engine [13]. The greatest developments in parabolic dish systems have taken place since the 70s through continuous research aimed at optimising their components [10], [14], [15]. This is demonstrated in a number of works dedicated to: the optimisation of the absorber temperature [16], [17], [18]; the influence of the solar collector design parameters on engine efficiency [19]; energy and exergetic analyses [20], [21] and the multi-objective optimisation [22], [23] of dish-Stirling systems. Moreover, the continuous advances in the design methods of these CSP systems, such those illustrated in [24], [25] and [26], have resulted in the realisation of numerous new demonstration plants all over the world [10].

Despite the continuous development of these systems, there have been several barriers which have not favoured their successful entry into the market of renewable electricity generation systems [10]. Some of these are typical of all CSP systems, i.e. the large initial capital investments and the subsequent uncertainty about the operation and maintenance costs [9], [27]. Others, are specific to dish-Stirling systems such as the difficulty in being integrated with storage systems compared to parabolic trough and linear Fresnel collectors [10]. On the other hand, several recent studies have shown that dish-Stirling systems could have numerous areas of application [15], such as: micro-cogeneration [11], potable water production [13], off-grid electrification and water pumping. Other interesting applications of dish-Stirling systems can certainly be developed in the future considering that this solar technology is very suitable for hybridisation with other renewable energy sources, according to a more general trend that began a few decades ago in this technological sector [28].

Another interesting study, shows that dish-Stirling systems have an environmental impact similar to that of photovoltaic systems, their main competitor, and that hybridisation with other renewable energy sources, such as bio-gas, could be the key factor in their commercialization [29]. Therefore, from a more general perspective of decarbonisation of electricity generation systems in the EU, national and international incentive policies will certainly be needed to encourage greater commercial penetration of this interesting typology of CSP system [27]. In order to implement these policies, it will be necessary to develop guidelines [30] which will have to be based on simple and realistic analysis tools to allow the evaluation of monthly and annual energy performance of these systems taking advantage, for example, of the hourly-based solar radiation databases usually available for each geographic location.

Among several numerical models used to evaluate the energy production of dish-Stirling systems based on real performance data from operating dish-Stirling plants [31], [32], [33], [34], the Stine model and its subsequent evolutions are the most widely used [30], [35]. These empirical models are based on the observation that there is usually a linear correlation between Direct Normal Irradiation (*DNI*) and the electrical power generated by dish-Stirling plants when the collector mirrors are clean [31], [32], [36]. The coefficients of the correlation can be calibrated using function test data on a full day with clear skies. These constants can be further corrected to take into account the effect of air temperature changes and degradation of optical efficiency due to soiling of the collector mirrors [30], [31], [32]. Above all, the effect of mirror soiling is a very important aspect in assessing the performance of a plant because, as experimental data show, it can lead to a significant reduction in the efficiency of the system. [36].

In this paper we show how data collected during the monitoring program of a demo dish-Stirling plant, recently built at a facility test site at Palermo University (Sicily), was used to assess experimentally the main factors affecting the performance of these systems. The plant site is characterised by a Mediterranean climate [37]. The tests performed on this plant are particularly interesting because there are few demonstrative dish-Stirling plants installed in the Central Mediterranean region and there is therefore no evidence of the effects of the particular micro-climate of this region on the performance and maintenance operations of this type of plant. This geographical area is, in fact, characterised by rather high levels of DNI [38] particularly during the warmer periods of the year. However, cloud decks associated with typical mid-latitude depressions and frontal passages are evident during cold periods of the year. In addition, desert dust transport events are frequent in this area. Dust transport occurs almost every day in different parts of the Mediterranean, especially during spring and summer [39], [40], [41], [42]. Both the frequent passage of clouds in the winter season and the dust deposition on collector mirrors in the spring and summer seasons can reduce the performance of dish-Stirling systems, yet these effects are often difficult to predict and manage. Using operating data measured on days when the concentrator mirrors were clean, we have elaborated a physical-numerical model based on a simplified energy balance of the collector (according to a well-established technique for modelling the CSP system performance [43]), taking into account the part-load efficiency curve of the Stirling engine [44]. Assuming further reasonable simplifying hypotheses, based on the experimental evidence, it has been possible to linearise the model obtaining a physical interpretation of the empirical coefficients of the model proposed by Stine. In addition, we compared the performance measured for the dish-Stirling system with soiled mirrors with that expected from the model with clean mirrors, for the same conditions of *DNI* and air temperature, using a technique proposed in the literature to assess the effect of module soiling on the energy production of photovoltaic systems [45]. Using this method, it was possible to define an average daily cleanliness index of the system for each day during the monitoring period. In a similar way to that proposed for photovoltaic systems [46], we compared the evolution of this index with the history of rains and dust deposition events recorded in the same period for the area where the plant is installed.

**2. Experimental setup and performance data analysis**

2.1 *The dish-Stirling demo plant in Palermo*

On 14 December 2017, the installation of a dish-Stirling plant at a facility test site at Department of Engineering of Palermo University was completed (see Fig. 1). The *CSP* unit, which was manufactured by the Swedish company *Ripasso Energy* and constructed by the Italian companies *Elettrocostruzioni S.r.l* and *HorizonFirm S.r.l.*, started operating in January 2018. During the following months, the demo plant underwent an intense test and monitoring program which was intended to assess the performance of the system and its possible commercialisation for the *CSP* Italian market. In the preceding years, *Ripasso Energy* had already installed a series of similar dish-Stirling units on a test site at Upington in the Republic of South Africa, which is a location characterised by very high levels of *DNI*.

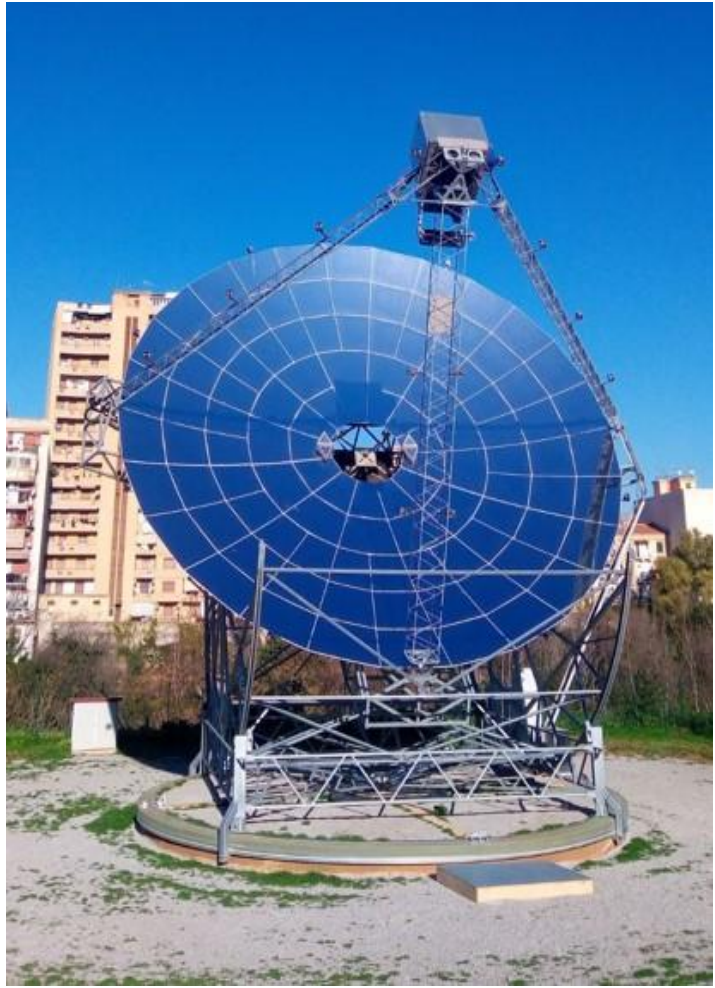

Figure 1. dish-Stirling unit at the Department of Engineering, University of Palermo, Italy.

The electric output of the *Ripasso CSP* unit is typically linearly proportional to the direct solar radiation with a net electric peak power of 31.5 kW$_e$ at a *DNI* value of 960 W/m$^2$. Nevertheless, in November 2012, the *Ripasso* South African plants set the current world record of energy conversion efficiency from solar-to-electric at 32%, making the *Ripasso* dish-Stirling a state-of-the-art system in the *CSP* sector [10]. The most relevant technical data on the *CSP* unit installed at the facility site at the University of Palermo has been summarised in Tab. 1 and is briefly described below.

Table 1: Technical data of the Ripasso Energy dish-Stirling unit.

| **Concentrator** | | |
|---|---|---|
| Dish diameter | 12 | [m] |
| Focal length | 7.45 | [m] |
| Effective aperture area | 101 | [m$^2$] |
| Geometric concentration ratio | 3217 | [ - ] |
| Reflectivity of clean mirrors | 95 | [%] |
| Total height during operation | 14 | [m] |
| Total weight | 8000 | [kg] |
| Occupation of soil area | 500 | [m$^2$] |
| **Power Conversion Unit** | | |
| Type of Stirling engine | 4 cylinders double acting | [ - ] |
| Displaced volume | 4 × 95 | [cm$^3$] |
| Typical output at *DNI*=960 W/m$^2$ | 31.5 @ 2300 rpm | [kW$_e$] |
| Receiver temperature | 720 | [°C] |
| Working gas | Hydrogen | [ - ] |
| Max gas pressure | 200 | [bar] |
| Power control | Pressure - speed control | [ - ] |
| Weight | 700 | [kg] |

The *Ripasso Energy* dish-Stirling unit consists of a paraboloidal concentrator, a Power Conversion Unit (PCU) and a tracking system. The paraboloidal dish reflector, with a diameter of about 12 m and an effective aperture area of 101 m$^2$, is made up of a lightweight, steel dish structure, mirror facets and three truss beams supporting the PCU. The facets, mounted on the steel structure, consist of sandwich panels with a low-iron, thin, back-silvered glass mirror surface characterised by high levels of reflectivity (95% for clean surfaces). The high precision dual-axis tracking system, on the other hand, is composed of the tracker structure, the actuators, the control cabinet and an annular concrete foundation occupying an area of about 500 m$^2$ and supporting the weight of the whole plant (about 8.7 tons). The tracker steel structure, in turn, is made up of two main parts: a carousel-type structure mounted on four wheels for azimuthal rotations and a cradle-type structure for elevation rotations. Finally, the PCU consists of a heat receiver, with an internal cavity, a Stirling engine, an alternator and a cooling system.

The tracking system, continuously following the daily path of the sun, allows the parabolic dish to collect, reflect off and concentrate the direct solar radiation onto the receiver of the PCU located at paraboloid's focal point (focal length of 7.45 m). The high precision tracking and optics systems permit a uniform distribution of the thermal energy on the receiver surface, which is adjacent to the hot side of the Stirling engine. Thus, the thermal energy absorbed by the receiver heats the working gas of the Stirling engine (hydrogen) to a temperature of about 720 °C. Part of this heat is firstly converted into mechanical energy by the engine and then transformed into three-phase electricity by the alternator connected to the crankshaft of the engine. The remaining heat is, instead, released into the environment by the fan cooling system mounted on the rear of the tracking structure and connected by a pipe and a heat exchanger to the coldest part of the Stirling engine. The cooling system, where water circulates, allows the maintenance of a high temperature difference between the hot and cold sides of the Stirling engine (720 °C and 70 °C, respectively). This is fundamental to guarantee high conversion efficiency from thermal to mechanical energy.

Although the *Ripasso CSP* unit is characterised by numerous innovative and highly efficient technological components, especially the tracking system and the receiver, the core part of the whole system remains the Stirling engine. The highly efficient *Ripasso* Stirling engine is an updated version of the original model (USAB 4-95) licensed from *Kockums AB* in 2008. Some data concerning the performance of the original USAB 4-95 Stirling engine, working with a temperature of 720 °C, can be found in [44]. In the past, this engine was installed on different dish-Stirling models, among which are: the MDAC systems located at different test facilities in the USA (1984-1985) [36] and the SES MPP systems (2008), one of which is located at the Sandia National Laboratories test site in New Mexico, USA [9]. These plants represent precursors to the evolved versions that *Ripasso Energy* has realised in recent years at the *Ripasso* test site at Upington, South Africa and at the facility site of the Department of Engineering of Palermo University, Italy.

2.2 *Measurement devices and data processing*

In the period between 11 January and 2 July 2018, an extensive measurement campaign was carried out to collect data related to the operation of the CSP plant located at the facility test site in Palermo. During this period, although the full operation of the solar plant was interrupted several times due to the maintenance and testing stages planned for the demo plant, it was possible to record, for 3,960 hours, the performance of the system under different conditions of solar irradiance, air temperature and cleanliness of the mirrors. The main physical quantities recorded during this stage were: the net electrical power output $\dot{E}_n$ using an energy meter (ABB - A43 212-100, with an accuracy of the active and reactive energy class of ±1% and ±2% respectively), the solar beam radiation $I_b$ using a pyrheliometer (Kipp&Zonen – SHP, with an average accuracy of ± 0.5%) and the external ambient temperature $T_{air}$ using a weather transmitter (Vaisala - WXT536 with an accuracy of ±0.3°C at +20°C). The other most relevant physical quantities measured in the same period were the parasitic electric power $\dot{E}_p$ and the limiting temperatures of the Stirling engine $T_h$ and $T_c$.

The original measured quantities were collected with a one-second temporal resolution in a MySQL database consisting of 14,256,000 records. The data was preliminarily filtered to avoid points representing mechanical transients. Thus, considering that the sampling frequency $f_s$ of the signals coming from the dish-Stirling plant is 1 Hz, it was first necessary to define a "moving average filter" procedure useful for both the reduction of the quantity of data and the filtering of the transients. The moving average is a low-pass filter, very common for regulating an array of sampled data, and is calculated according to the following formula [47]:

$$y(i) = \frac{1}{N}\sum_{j=0}^{N-1} x(j) \qquad (1)$$

where $x(j)$ is the monitored input signal, $y(i)$ is the resulted signal, and $N$ is the considered number of points. In the present case study, $i$ is the index referring to minutes, $j$ is the index referring to the seconds of the considered minute, and $N$ is equal to 60. This kind of filter is optimal for many common problems since it reduces random white noise while keeping the sharpest step response. An example of a moving average filter application is shown in Fig. 2, where the previously described procedure has been applied to a squared signal affected by random white noise.

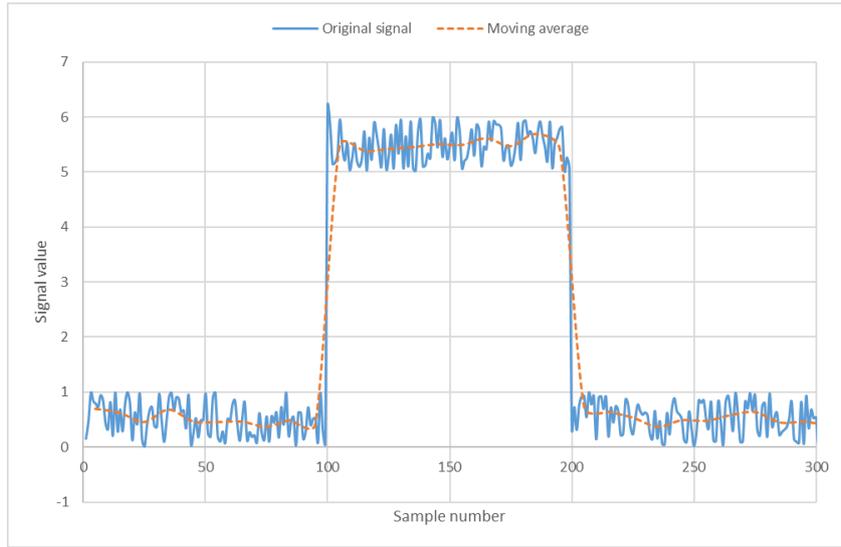

Fig. 2: comparison between an original square signal and a filtered signal expressed by Eqn. (1) over a time series of 300 seconds with 1 Hz of sampling frequency.

On average, the amount of noise reduction is equal to the square root of the number of points. In the present case study, a 60 point moving average filter reduces the noise by a factor of 7.7. Thus, after defining a cut-off frequency $f_c$ for which the attenuation of the output is larger than 3 dB, this frequency is equal to [47]:

$$f_c \approx \frac{0.442947}{\sqrt{N^2-1}} \cdot f_s \approx \frac{0.443}{N} \cdot f_s = 0.00738\, Hz \qquad (2)$$

Furthermore, all the data for which the revolutions per minute of the Stirling engine are over 2020 has been filtered out because this is indicative of a transient condition. As a result of the filtering procedures, it was possible to select the data from phases where the system worked continuously and in stationary conditions. This data was used for the analysis presented in this study.

2.3 *Analysis and discussion of the experimental results*

Preliminary to the analysis, the experimental data collected during the testing campaign (occurred between 11 January 2018 and 2 July 2018) was divided into two groups classified in accordance with whether they were recorded on days when the collector mirrors were clean or not. In fact, during the period in which the experimental data was recorded, the dish-Stirling plant was subject to numerous rainfall cycles of different intensity and sirocco events with associated variable levels of dust deposition. For this reason, we assumed that most of the analysed records correspond to conditions in which the collector mirrors were naturally characterised by different (unknown) levels of soiling. On the other hand, (although no reflectometric measurements were carried out during this preliminary phase of the experiment) a total of 13 days were selected from the total data set in which the mirrors were classified as clean also in relation to the special conditions occurring in the preceding days: the first 4 days between 18 January and 2 February immediately after removing the protective films from the mirrors and in a period in which the frequent rains kept them clean; another day on 30 May after a mirror washing had taken place on 23 May; and finally 8 other days between 16 June and 2 July when the mirrors were particularly clean as a result of heavy rains occurring on 14-15 June. On most of the other days, with the exception of these 13, the effect of dust deposition on the surface of mirrors was often observed even by the naked eye (by comparing the mirrors mounted on the collector with a perfectly clean sample mirror), such as on 6 March, as a result of a significant Saharan dust transport event (Sirocco) occurring from 28 February to 5 March.

With this in mind, we decided to follow the approach suggested by Sandia National Laboratories for the analysis of the performance of dish-Stirling plants [32]. The data of net electric power, corrected for the effect of the ambient temperature, was correlated with the corresponding values of direct solar irradiation and plotted in Fig. 3. In relation to the correction factor for the air temperature, it should be noted that in the first phase of the testing stage (between January and February) the average outdoor temperature ranged between 15-16 °C while in the last period (between May and July) between 26-28 °C.

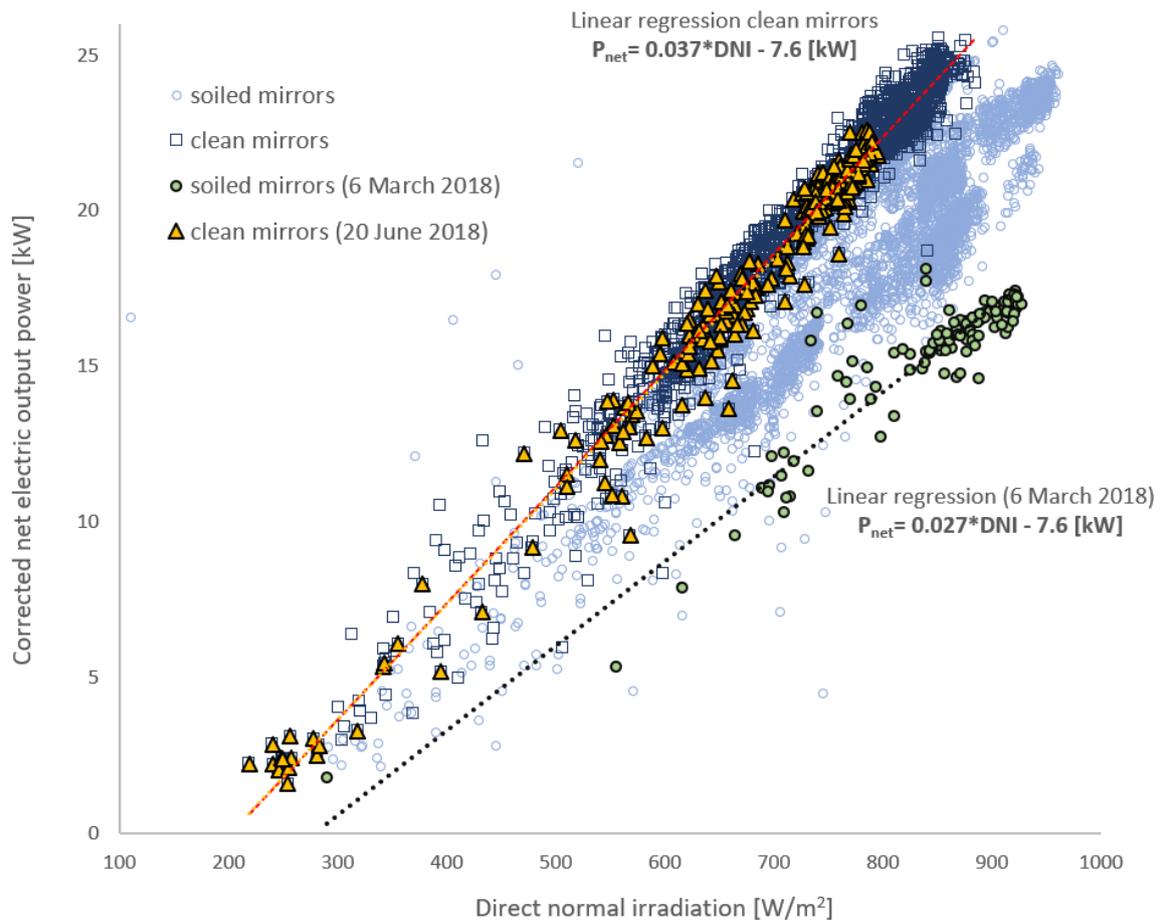

Fig. 3: Corrected net electric output power [32] versus *DNI* for the Palermo dish-Stirling unit (the records have been categorised corresponding to clean or soiled mirrors).

The data of air temperature, *DNI* and electrical power output for a day when the mirrors were clean (20 June 2018) are plotted in Fig. 4 as a function of time. The same data is also highlighted with yellow triangles in Fig. 3. On this day the air temperature was almost constant at about 27 °C, the *DNI* had a wide variation between 320-800 W/m$^2$, while the corresponding net electrical output power ranged between 4 kW and 22 kW. This day, as shown in Fig. 4, was particularly characterised by scattered variations of the *DNI* levels since a first slow passage of clouds occurred between 12:00 PM and 12:45 PM and a second rapid sequence of cloud transitions started at around 2:00 PM and concluded with a thunderstorm at about 3:00 PM.

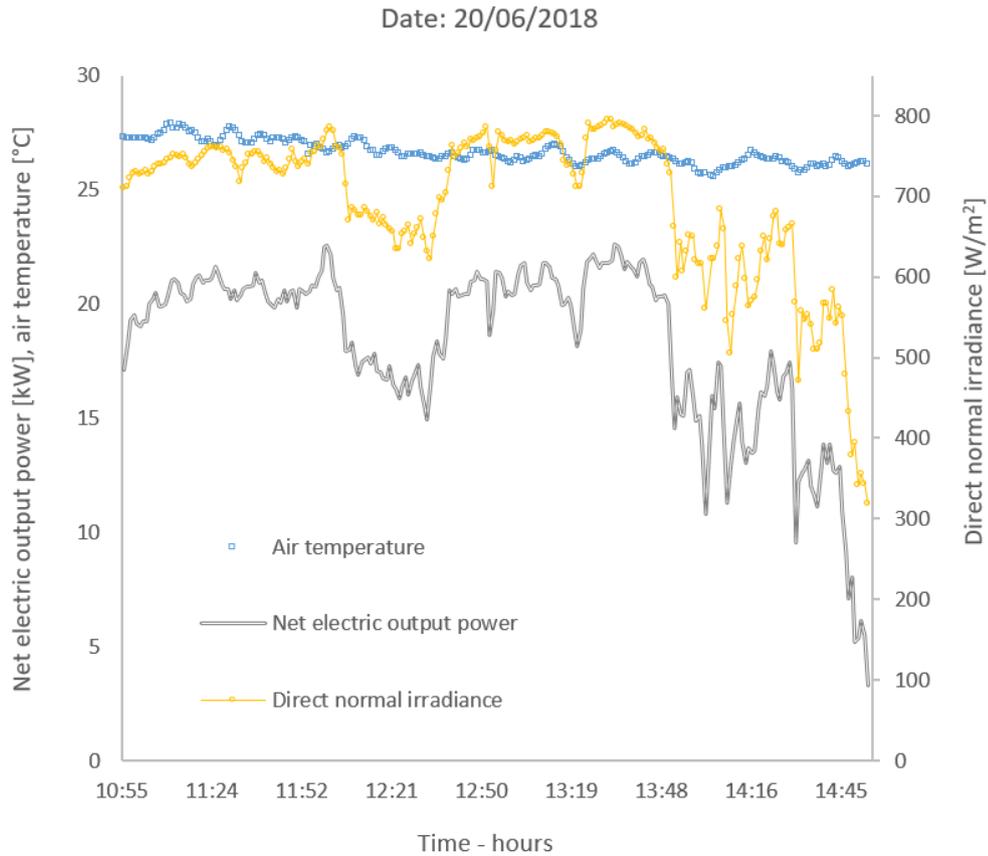

Fig. 4: A comparison of *DNI* and net electric output power as a function of time, alongside variation of air temperature, measured on 20 June 2018.

Thus, from the analysis of the resulting data plotted in Fig. 3 and Fig. 4 it is possible to observe that:

- The selected data corresponding to clean mirrors is well-correlated by a linear relation ($R^2$=0.96) representing the upper limit of the experimental point cloud
- all the other data shows that the plant performance was lower compared to that of the first group with similar values of *DNI*
- the data recorded on the 6 March (the day when the mirrors were visibly soiled after a Sirocco event) represents the lower limit of the experimental point cloud
- the data recorded on 20 June is well-aligned with that of clean mirror conditions despite the scattered variations in the *DNI* levels due to the fast passage of clouds on this day.

Therefore, considering both the methods used to select the clean mirror data and the previous observations, it can be deduced that a large majority of the reported production losses during the testing campaign were caused by the soiling of the mirrors and only marginally by other factors such as: sun tracking errors, fast passage of clouds, engine transients, etc.

With more specific regard to the clean mirror performance of the dish-Stirling plant in Palermo, the correct net electrical output power data was correlated with the corresponding *DNI* value s by means of a linear relationship, as suggested by the present experimental data and in accordance with the model proposed by Sandia National Laboratories for the SES system [32]. The equation in question has been reported in Fig. 3. Using this equation with $I_b$ =960 W/m² it is possible to extrapolate a net electric output power equal to $\dot{E}_n$=28 kW, which added to the average value of the parasitic electric consumption recorded during the operation of the Palermo system ($\dot{E}_p$=3.6 kW), makes it possible to calculate a gross electric output power equal to $\dot{E}_p$=31.6 kW. The latter value is very close to the rated gross electrical power output ($\dot{E}_p$=32.6 kW) that can be calculated from the data declared by the manufacturer (see Tab. 1) for the dish-Stirling system with clean mirrors ($\dot{E}_n$=31.5 kW and $\dot{E}_p$=1.1 kW). This consistency confirms that the peak performance of the plant is in line with what was expected for this particular system and that the selection of the set of data corresponding to the condition of clean mirrors was apt. On the other hand, the discrepancy between the parasitic electric power consumption values indicates that it will be necessary to work further on the optimisation of the electric consumption of both the sun tracking and cooling systems of the demo plant in Palermo.

However, as already illustrated in Fig. 3, the effect of the mirror soiling can produce much greater reductions in performance than those due to parasitic consumptions, especially at higher levels of *DNI*. In order to quantify these effects on the optical efficiency of the system, we used the method proposed by Stine [31] through the introduction of a reductive factor to the slope of the linear function of Fig. 3. Therefore, by fitting with a linear expression the data of $\dot{E}_n$ as a function of *DNI* relative to the measurements from 6 March, the day when the reflectivity of the mirrors reached its minimum, it was possible to deduce a reduction in the optical efficiency of the collector equal to about 30%, calculated by comparing the slope of the linear expression corresponding to clean mirrors with that of soiled mirrors (see Fig. 3).

Finally, Fig. 5 reports the data of the receiver temperature as a function of the *DNI* variations all of which were recorded during the experimental campaign (with cleaned and soiled collector mirrors). Thus, analysing the complete data set, it is possible to observe that the data varies between a minimum of about $T_r$= 710 °C and a maximum of $T_r$=728 °C with an average of $T_r$=719.79 °C. However, in about 85% of the data the temperature is within the narrow range of $T_r$= 720±0.5 °C. The remaining values represent conditions in which there have been small transients caused, for example, by the start-up of the system or the rapid passage of cloudy formations, etc. These variations with respect to the average, as can be seen from Fig. 5, are independent of the DNI values. Thus, observing this data it can be deduced that, regardless of the conditions of irradiation and ambient temperature, the receiver temperature remained almost constant at about $T_r$=720 °C. This observation, as will be clear later on in the discussion presented in this article, will be fundamental for the development of the numerical model presented in this work.

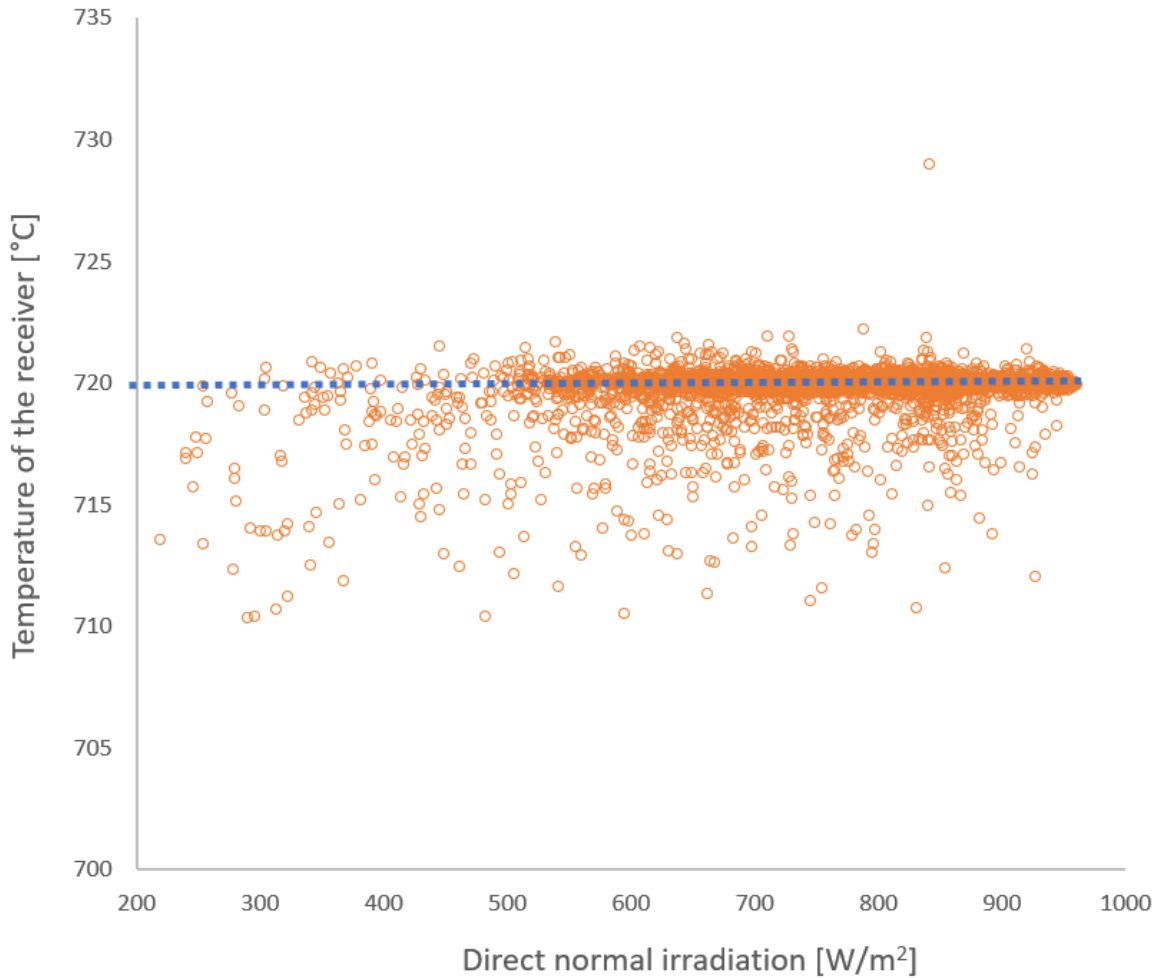

Fig. 5: Temperature of the receiver as a function of the variations of *DNI* recorded during the testing campaign. The points represent data when the collector mirrors have different levels of soiling.

## 3. Theoretical analysis

### 3.1 *Energy balance of the dish-Stirling system*

In dish-Stirling systems (see Fig. 6), the mirrors constituting the primary optics of the collector (usually with a parabolic shape) concentrate the sun's light on the focal point of the collector, where a cavity-type receiver absorbs and transfers the concentrated solar energy to a Stirling engine driving an electric generator. Unlike other types of solar collectors (photovoltaic and thermal), CSP systems, such as solar dish-Stirling plants, can only transform the direct component of solar radiation into thermal/electric power. Therefore, the rate of solar energy reflected off the mirrors of a collector continuously tracking the sun can be calculated as [16], [21], [43]:

$$\dot{Q}_{sun} = I_b \cdot A_n \tag{3}$$

where $I_b$ is the solar beam radiation and $A_n$ is the net effective surface of the mirrors (projected mirror area) of the dish collector. The latter quantity can be calculated from the total collector aperture area $A_n$ by:

$$A_n = \eta_m \cdot A_a \tag{4}$$

where $\eta_m$ is a factor taking into account all of the causes of the losses of effective reflective area, such as blocking and shading, and the gap between the mirror facets [44]. However, not all of this solar power is effectively reflected off the mirrors, intercepted by the cavity receiver aperture and then absorbed by its surface, because of different factors that affect the optical performances of the collector, the main ones being [44]:

- reflectance of the mirrors
- mirror surface slope errors
- soiling of the mirrors
- sun tracking system errors
- receiver positioning errors (often combined with mirror errors)
- wind-induced vibrations
- effective absorbance of the cavity receiver
- water condensation at the mirror surface during the early hours of the day

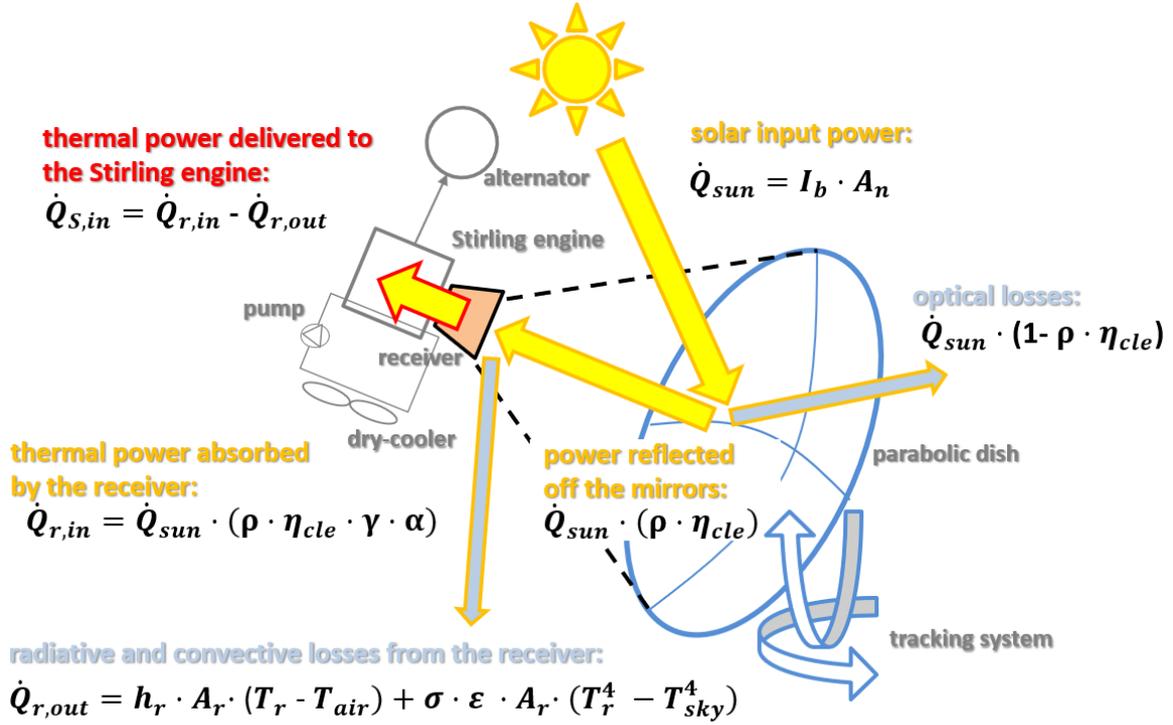

Figure 6. Schematic diagram showing the energy balance of the solar collector.

Taking into account a combination of these effects, the concentrated solar power effectively absorbed by the cavity receiver is usually expressed as:

$$\dot{Q}_{r,in} = \dot{Q}_{sun} \cdot \eta_o \qquad (5)$$

where $\eta_o$ is the optical efficiency of the concentrator system, defined as:

$$\eta_o = \rho \cdot \eta_{cle} \cdot \gamma \cdot \alpha \qquad (6)$$

where $\rho$ is the clean mirror reflectance (which also incorporates the losses due to slope mirror errors), $\eta_{cle}$ is a factor which takes into account the mirror cleanliness, $\gamma$ is the intercept factor of the collector and $\alpha$ is the effective cavity receiver absorptance [48].

In other words, optical efficiency $\eta_o$ can be interpreted as the ratio between solar power absorbed by the receiver $\dot{Q}_{r,in}$ and the rate of solar power hitting the collector, while the intercept factor $\gamma$ is the ratio between the rate of energy hitting the receiver $\dot{Q}_{sun}$ and the effective concentrated solar power reflected off the mirrors of the concentrator collector. The intercept factor γ, in particular, incorporates the combination of errors associated with the control of sun tracking, the erroneous positioning of the receiver and defocusing, due to wind-induced vibrations of the collector structures.

It is also necessary to consider the additional convective and radiative thermal losses from the aperture of the cavity receiver, due to the high temperatures reached at its internal surface. Then the total power losses from the receiver can be defined as the following sum:

$$\dot{Q}_{r,out} = \dot{Q}_{con} + \dot{Q}_{rad} \tag{7}$$

where $\dot{Q}_{con}$ and $\dot{Q}_{rad}$ are the convective and radiative losses respectively. The convective losses, can be expressed as [16], [18], [49]:

$$\dot{Q}_{con} = h_r \cdot A_r \cdot \left( T_r - T_{air} \right) \tag{8}$$

where $h_r$ is the effective convective coefficient of the cavity receiver, $A_r$ is the surface area of the aperture of the receiver and $T_r$ and $T_{air}$ are the internal receiver and external ambient temperatures, respectively. The second component of thermal losses, the radiative losses, are generally modelled as [18], [49]:

$$\dot{Q}_{rad} = \sigma \cdot \varepsilon \cdot A_r \cdot \left( T_r^4 - T_{sky}^4 \right) \tag{9}$$

where $\varepsilon$ is the effective emissivity of the cavity receiver [48], [50], $\sigma$ is the Stefan–Boltzmann constant and $T_{sky}$ is the effective sky temperature. This temperature depends on the atmospheric conditions and is usually related to the air temperature by empirical formulas like the following [16], [18], [19]:

$$T_{sky} = 0.0552 \cdot T_{air}^{1.5} \tag{10}$$

It is not trivial to observe that the greater the aperture of the cavity receiver $A_r$, the greater the thermal losses $\dot{Q}_{r,out}$ and the higher the intercept factor γ [44]. Thus, the correct design of the dimension of the receiver aperture should always be based on the trade-off between two opposing tendencies [44]: the reduction of thermal loss from the receiver and the increase in optical efficiency, the latter being proportional to the intercept factor as shown in Eq. 6.

From the energy balance of the receiver, it is possible to express the thermal power which is effectively delivered to the Stirling engine by the following difference [16], [19], [21]:

$$\dot{Q}_{S,in} = \dot{Q}_{r,in} - \dot{Q}_{r,out} \qquad (11)$$

which, using all of the above equations, can be recast as [7], [21], [51]:

$$\dot{Q}_{S,in} = I_b \cdot A_n \cdot \eta_o - A_r \cdot \left[ h_r \cdot \left( T_r - T_{air} \right) + \sigma \cdot \varepsilon \cdot \left( T_r^4 - T_{sky}^4 \right) \right] \qquad (12)$$

Thus, dividing Eq. 12 by Eq. 3, it is possible to make explicit the expression describing the thermal efficiency of the solar collector as [18], [22], [51]:

$$\eta_c = \frac{\dot{Q}_{S,in}}{\dot{Q}_{sun}} = \eta_o - \frac{1}{I_b \cdot C_g} \cdot \left[ h_r \cdot \left( T_r - T_{air} \right) + \sigma \cdot \varepsilon \cdot \left( T_r^4 - T_{sky}^4 \right) \right] \qquad (13)$$

where $C_g$ is the solar collector geometric concentration ratio defined as:

$$C_g = \frac{A_n}{A_r}. \qquad (14)$$

After calculating the thermal solar input power to the CSP system, the amount of this power that can be transformed into electrical power $\dot{E}_n$ needs to be determined, but it is first necessary to evaluate the mechanical output power of the Stirling engine by exploiting the energy balance of the PCU (see Fig. 7).

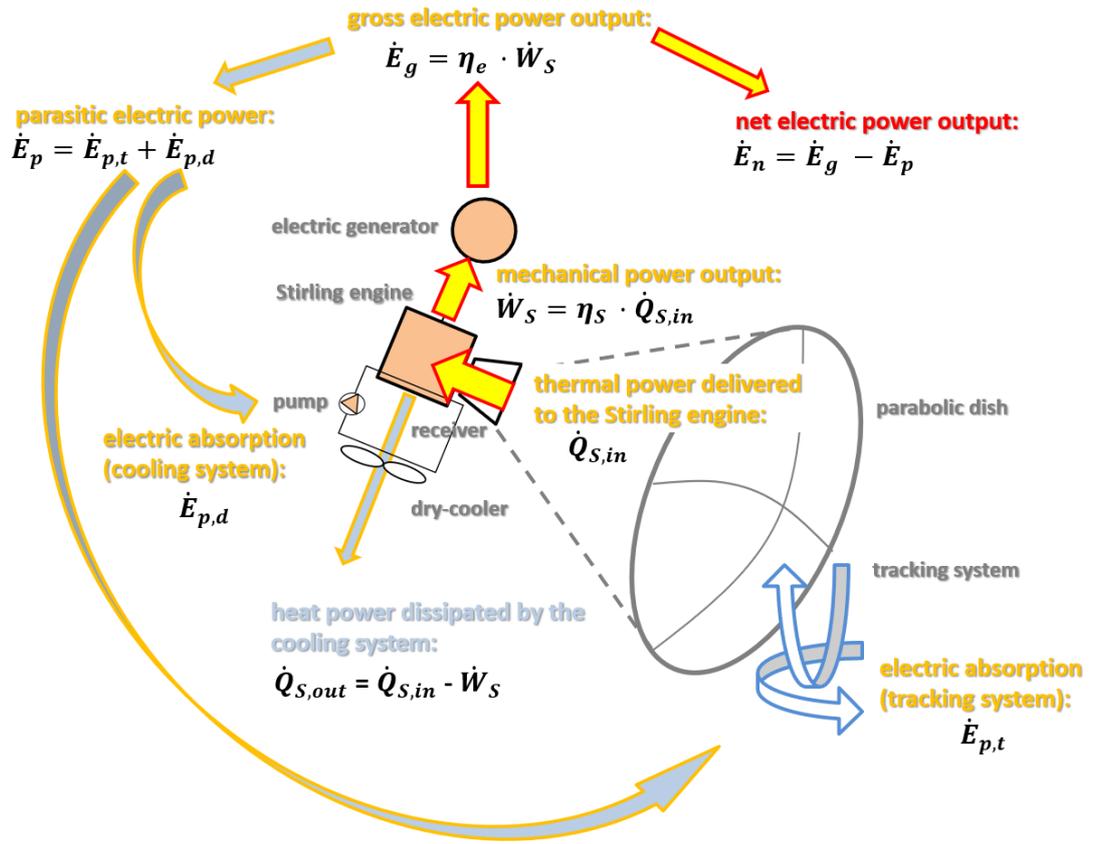

Figure 7. Schematic diagram showing the energy balance of the *PCU*.

The mechanical output power of the Stirling engine is related to the input thermal power by the efficiency of the engine itself $\eta_s$ as:

$$\dot{W}_S = \eta_S \cdot \dot{Q}_{S,in} \quad (15)$$

From a theoretical point of view, the Stirling cycle consists of four internally reversible processes: two processes are isothermal compressions at temperature $T_h$ and $T_c$; the other two thermodynamic processes are isochoric (constant volume). During isothermal expansion at temperature $T_h$, heat is added to the working fluid, while during isothermal compression at temperature $T_c$, heat is rejected to the surroundings. A regenerator with full effectiveness allows the heat rejected during the isochoric rarefaction to be transferred during the isochoric compression. This Stirling cycle exhibits the same Carnot thermal efficiency.

The efficiency of the Stirling engine, in turn, could be generally related to the reversible Carnot efficiency of the engine $\eta_{S,C}$, between the same $T_h$ and $T_c$ limiting temperatures, as [12], [18], [51]:

$$\eta_S = \eta_{S,ex} \cdot \eta_{S,C} \tag{16}$$

where $\eta_{S,ex}$ is the exergetic efficiency of the engine. The Carnot cycle efficiency, which is a function of both the heat input temperature $T_h$ and the heat rejection temperature $T_c$ [13], [17], [18]:

$$\eta_{S,C} = \left(1 - \frac{T_h}{T_c}\right) \tag{17}$$

represents the ratio between the maximum theoretical mechanic power output of the ideal Stirling cycle operating between the limiting temperatures $T_h$ and $T_c$, and the actual thermal power input $\dot{Q}_{S,in}$ The exergetic efficiency $\eta_{S,ex}$, represents the ratio between the actual mechanical power output $\dot{W}_S$ and the maximum mechanical power. The value of the exergetic efficiency takes into account the irreversibilities of the engine [20] and is found between $\eta_{S,ex}$ = 0.55-0.88 [12], [18]. More generally, it is also necessary to consider that the performance of these kinds of systems depends on many factors [52], among which is the effect of the partialisation of the input thermal power due to fluctuations in solar radiation [30]. For solar dish-Stirling engines, it is often assumed $\eta_{S,ex}$ = 0.5 [13], [18], though there are relatively few experimental works on the performance of these engines.

In addition, using the energy balance of the Stirling engine (see Fig. 7), based on the assumption that the heat delivered to the Stirling engine and the heat dissipated $\dot{Q}_{S,out}$ occur simultaneously [7], it is also possible to express the thermal power that must be dissipated by the cooling systems as:

$$\dot{Q}_{S,out} = \dot{Q}_{S,in} - \dot{W}_S = \left(1 - \eta_s\right) \cdot \dot{Q}_{S,in} \tag{18}$$

Based both on the analysis of the real data of partial load operation of the original USAB 4-95 Stirling engine (working with $T_h$ = 720 °C) [44] and on the performance of the Stirling engine of the SES plant [35], we decided to approximate with a linear regression

the curve between the thermal input power and the mechanical output power of the engine as:

$$\dot{W}_S = (a_1 \cdot \dot{Q}_{S,in} - a_2) \cdot R_T \tag{19}$$

where $a_1$ and $a_2$ are the two positive fitting constants and $R_T$ is the ratio of a reference temperature $T_o$ to the current ambient temperature $T_{air}$ (both expressed as Kelvin degrees):

$$R_T = \frac{T_o}{T_{air}} \tag{20}$$

This last correction factor was introduced, inspired by what was proposed in the Stine model [30], [31] and similar ones [32], to take into account the effect of variations in outside temperature on the efficiency of the Stirling engine. Finally, combining Eq. 15 with Eq. 19, it follows:

$$\eta_S = \left( a_1 - \frac{a_2}{\dot{Q}_{S,in}} \right) \cdot R_T \tag{21}$$

which is an expression of the efficiency of the Stirling engine as a function of the partialisation of thermal input power and the ambient temperature.

Using the mechanical output power of the engine, it is then possible to evaluate the gross electric power of *PCU* by:

$$\dot{E}_g = \eta_e \cdot \dot{W}_S \tag{22}$$

where $\eta_e$ is the electrical efficiency of the alternator. Finally, the net electric power of the *CSP* system can be determined by subtracting the total parasitic electric absorption $\dot{E}_p$ from the electric gross power:

$$\dot{E}_n = \dot{E}_g - \dot{E}_p \tag{23}$$

The total parasitic power of the dish-Stirling unit, in turn, is mostly the sum of two contributors: the electric absorption of the control system and tracking motors $\dot{E}_{p,t}$ and those of the pumps and fans of the cooling system $\dot{E}_{p,d}$, i.e.:

$$\dot{E}_p = \dot{E}_{p,t} + \dot{E}_{p,d} \tag{24}$$

Finally, the instantaneous efficiency from solar-to-electric for all the dish-Stirling plants can be calculated as the ratio between the net electrical power produced and the solar power hitting the concentrator mirrors:

$$\eta_{DS} = \frac{\dot{E}_n}{\dot{Q}_{sun}}. \tag{25}$$

By substituting: the right-hand-side of Eq. 12 into Eq. 19 and, then the right-hand-side of the obtained expression into Eq. 22 and, finally, the resulting expression into Eq. 23, it is possible to obtain, in a compact form, the following equation:

$$\dot{E}_n = \eta_e \cdot \eta_o \cdot a_1 \cdot A_n \cdot R_T \cdot I_b - \left[ \eta_e \cdot a_1 \cdot \dot{Q}_{r,out} + a_2 \cdot R_T + \dot{E}_p \right] \tag{26}$$

where $\dot{Q}_{r,out}$ are the thermal losses from the receiver described by Eqs. 7-10. The model described by Eq. 26, once calibrated with the real operating data, can be applied in order to predict the net electric power output of the dish-Stirling system taking into account:

- the changes of the direct normal irradiation and external temperature
- the optical efficiency (by the parameters of Eq. 6)
- the effect of mirror soiling (by the cleanliness index in Eq. 6)
- the thermal losses from the receiver (by the parameters of Eqs. 7-10) assuming the receiver temperature is constant during the operation of the plant
- the expression of the thermal efficiency of the Stirling engine as a function of the partialisation of the thermal input loads (by Eq. 19) and external air temperature (by Eq. 20)
- the parasitic powers of the tracking and cooling systems of the PCU (by Eq. 24)

In Fig. 8, the calculation flowchart describing the calibration procedure of the model is presented. The procedure is essentially aimed at the estimation of the two slope $a_1$ and the intercept $a_2$ of the linear relationship between $\dot{W}_S$ and $\dot{Q}_{S,in}$. This algorithm must be performed using the dataset corresponding to the condition of clean mirrors of the collector as input. Therefore, in the described procedure it has been assumed that the $\eta_{cle}$ coefficient is always unitary. Finally, Fig. 9, represents the flowchart describing the calculus of the net electric output power from the plant using the model with its parameters already calibrated.

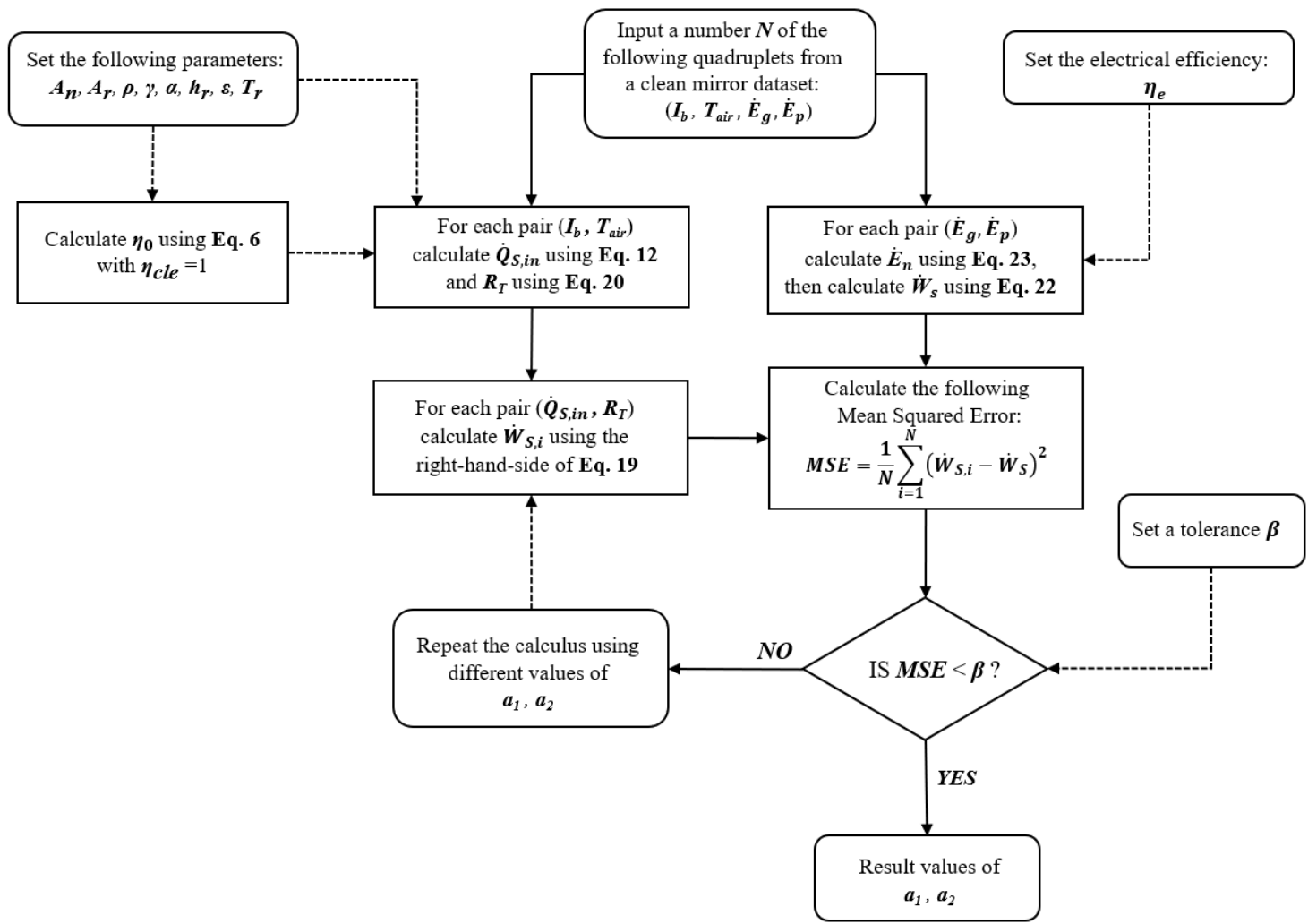

Figure 8. Model calibration procedure flowchart using a dataset corresponding to the condition of clean mirrors of the collector.

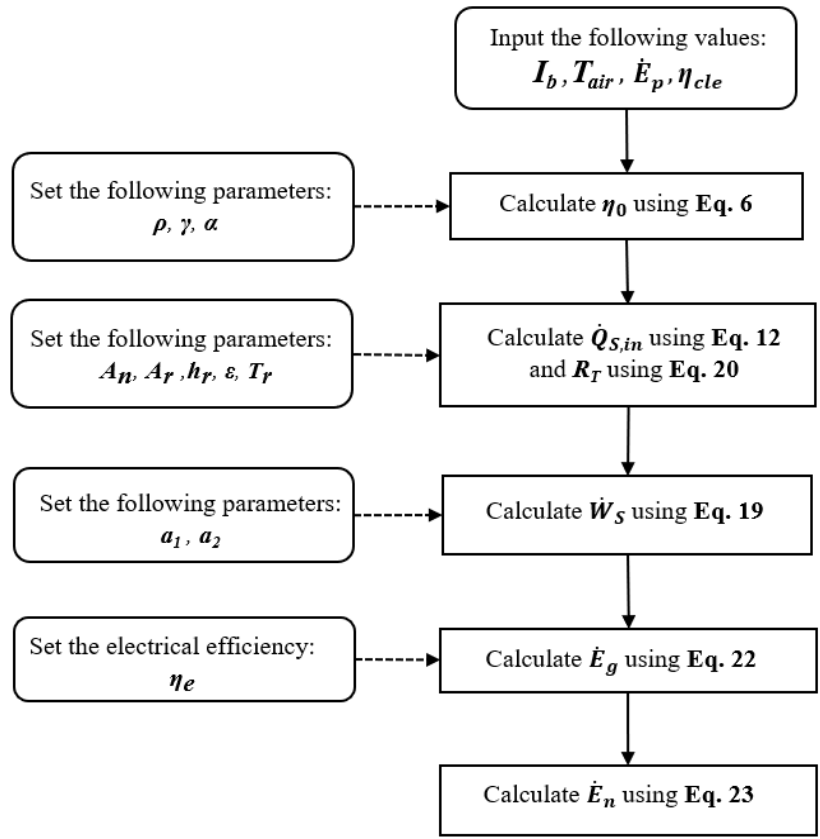

Figure 9. Calculation flowcharts for the determination of the net electric output power.

3.2 *A linear model of dish-Stirling electric power generation*

From the inspection of Eqs. 7-10, using an appropriate set of parameters, it may easily be verified that, if the receiver temperature $T_r$ is kept almost constant during plant operation (as, for example, observed in the measurements dataset of the dish-Stirling in Palermo), the thermal losses from the receiver are not very sensitive to variations in $T_{air}$. Under this assumption, it follows from Eq. 26, that the gross electric power output can be expressed as:

$$\dot{E}_g = \eta_e \cdot \eta_o \cdot a_1 \cdot A_n \cdot I_b - \left[ \eta_e \cdot a_1 \cdot \dot{Q}_{r,\text{out}}^{ave} + a_2 \right] \cdot R_T \qquad (27)$$

where $\dot{Q}_{r,out}^{ave}$ is an average value of the thermal losses from the receiver. Considering, moreover, a constant value of the cleanliness index of the mirrors ($\eta_o$ constant from Eq. 6) and an average value of the air temperature ($R_T$ constant from Eq. 20), it can be deduced that Eq. 27 simplifies to a linear relationship between the gross electrical power output and the solar beam radiation. Interestingly, this linearity emerges from the analysis of the operational data of existing dish-Stirling plants and is the basis of the empirical model proposed and validated by Stine [30], [31].

As in the Stine model, Eq. 27 shows that the optical efficiency reduction due to the soiling of mirrors has the practical effect of changing the slope of the linear equation [30], [31]. Stine in particular introduces this effect through the *corrected insolation* concept defined as the product of the current beam radiation with a reflectance ratio, the latter defined as the ratio between the measured current reflectance (with soiled mirrors) and that with clean mirrors [31]. On the other hand, unlike the original Stine model, we have defined the correction term $R_T$ in Eq. 26 in analogy to what is proposed in the Sandia National Laboratories model for SES dish-Stirling plants [32], i.e., as a ratio between a rated temperature and the current air temperature (see Eq. 20).

Moreover, if we further assume a constant value for the parasitic absorptions, equal to the average of the values measured during the operation of the plant $\dot{E}_p^{ave}$, we can obtain from Eq. 27:

$$\dot{E}_n = \eta_e \cdot \eta_o \cdot a_1 \cdot A_n \cdot R_T \cdot I_b - \left[ \eta_e \cdot a_1 \cdot \dot{Q}_{r,\text{out}}^{ave} + a_2 \cdot R_T + \dot{E}_p^{ave} \right] \qquad (28)$$

which represents a linear expression between the net output power generated by the plant and the normal direct radiation hitting the collector mirrors. Also this linearity has been extensively observed by the experimental performance data of numerous dish-Stirling plants operating with clean mirrors [9], [32], [36] and is further confirmed by the results from the Palermo plant which are presented in this work.

Thus, the analytical expressions represented by Eq. 27 and Eq. 28 can be considered to be the main result of this work as they provide a theoretical interpretation of the empirical models presented by Stine for the SBP dish-Stirling and by Sandia National Laboratories for the SES dish-Stirling. The latter models, calibrated on a limited number of tests carried out on pilot plants, have been used successfully to extrapolate the energy production and economic performance of dish-Stirling plants for potential commercial sites. The main advantage to using the new model presented in this paper, is that it can also be used to carry out technical-economic optimisation analyses of dish-Stirling systems such as those presented, for example, by Sandia National Laboratories for SES plants [35]. Another possible application of the proposed approach is to re-analyse the dish-Stirling performance data already available in the literature [9], with the purpose of carrying out a comparative analysis of the performance of these systems. For example, it is possible to extrapolate the efficiency curves of the engines with which they are equipped using the system waterfall charts and data on net electric power output as a function of *DNI*.

To this end, we can further simplify Eq. 28 assuming an average value of the ambient temperature $T_{air}^{ave}$ and, therefore, an average value of the corrector factor $R_T^{ave}$, obtaining:

$$\dot{E}_n = b_1 \cdot I_b - b_2 \tag{29}$$

where the slope is equal to

$$b_1 = \eta_e \cdot \eta_o \cdot a_1 \cdot A_n \cdot R_T^{ave} \cdot 10^{-3} \tag{30}$$

and the intercept is equal to

$$b_2 = \eta_e \cdot \left( a_1 \cdot \dot{Q}_{r,out}^{ave} + a_2 \cdot R_T^{ave} \right) + \dot{E}_p^{ave} \tag{31}$$

In Eq. 30 and Eq.31 we have considered $b_1$ expressed in [kW·m$^2$/W] and $b_2$, $a_2$, $\dot{Q}_{r,out}^{ave}$ and $\dot{E}_p^{ave}$ in [kW]. The parameters $b_1$ and $b_2$ in Eq. 29 can be estimated through a linear regression of the data of $\dot{E}_n$ as a function of the *DNI*. Then, using the waterfall charts

(elaborated for the peak performance of the collector), it is possible to deduce: the reflectance of the mirror $\rho$, the intercept factor of the collector $\gamma$ and the peak value of the thermal input power to the Stirling engine $\dot{Q}_{S,in}^{max}$. Thus, by estimating a realistic value of the effective absorbance of the receiver $\alpha$, it is then possible to calculate the average value of the thermal losses from the receiver from the following expression:

$$\dot{Q}_{r,out}^{ave} = \alpha \cdot \rho \cdot \gamma \cdot I_b^{max} - Q_{S,in}^{max} \tag{32}$$

where $\dot{I}_b^{max}$ indicates the maximum design input value of *DNI*. Finally, from Eq. 30, it is possible to calculate:

$$a_1 = 10^3 \cdot \frac{b_1}{\eta_e \cdot \eta_o \cdot A_n \cdot R_T^{ave}} \tag{33}$$

and, after introducing Eq. 33 in Eq. 31:

$$a_2 = \frac{\left[ b_2 - \dot{E}_p^{ave} \cdot \eta_o \cdot A_n - 10^3 \cdot b_1 \cdot \dot{Q}_{r,out}^{ave} \right]}{\eta_e \cdot \eta_o \cdot A_n \cdot R_T^{ave}} \tag{34}$$

where $a_1$ and $a_2$ are the slope and the intercept of the Eq. 19, respectively.

### 3.3 *A method for estimating the daily averaged cleanliness index of mirrors*

Every solar concentrating device like dish-Stirling system is affected significantly by the presence of dust in two ways: one is the reduction of incoming solar radiation, and the other is the deposition of dust particles on the surface of the reflectors. The accumulation of soiling on the mirrors reduces the reflectivity and can result in an uneven energy flux on the receiver [36]. In other words, progressive soiling leads to an increasing reduction in the optical efficiency of the collector and, therefore, to a decrease in the electrical power generated by the system. When production losses are no longer economically acceptable, it is, therefore, necessary to wash the mirrors. The frequency of washing cycles, however, is clearly dependent on the particular climatic conditions of the site where the plant is installed.

In the Mediterranean region, in particular, dust greatly influences radiative transfer. The radiation can be attenuated by a few tens of W/m$^2$, and during severe episodes even by a few hundred W/m$^2$. The deposition of dust is quite significant. It has been estimated, with experimental and modelling tools, that approximately 108 metric tons of dust are deposited annually over Mediterranean waters, and almost the same amount is transferred towards the landmass of Europe [53], [54].

The main source of material for soiling are the various types of aerosols in the atmosphere. The aerosols are of natural and/or anthropogenic origin [55], [56], [57]. Anthropogenic particles are regularly released by human activities (mainly combustion) or the by-product of conversion of gaseous pollutants to aerosols mainly sulphates and nitrates [55], [57], [58]. They are of submicron size and are highly hygroscopic. Deposition of these aerosol particles on the mirrors occurs mainly during high moisture conditions (e.g. during night hours or when drizzling). The main source of natural particles in the Mediterranean region is Saharan dust. Saharan dust transport exhibits seasonal variability with a peak during springtime [41], [55], [57]. The dust particle size is from submicron to about 50 μm. The transport is of an episodic type, with each episode lasting from one to three days regularly. These can be predicted relatively easily by utilizing Numerical Weather Prediction (NWP) models with a coupled dust cycle sub-model (e.g. SKIRON/Dust system, [59]).

Such systems have been developed for the Mediterranean region in the 2000s and are regularly in operation by providing 3-5 day predictions of dust concentration and deposition (dry and wet). Other sources of naturally-produced aerosols are sea-salt and pollen from trees. Sea salt is the remnant of seawater droplet evaporation (wave breaking, bubble generation and breaking). Sea salt is highly hygroscopic and sticks to the mirror surfaces relatively easily. Sea-salt regularly affects coastal regions [60]. Pollen is a more hydrophobic type of particle that does not stick easily to the mirrors.

The assessment of energy production losses due to soiling of the collector mirrors is an important part of the performance monitoring of a *CSP* plant. The soiling level is generally estimated by periodically measuring the average reflectivity of the collector mirrors [36]. Experimental data describing the history of reflectivity for a dish-Stirling plant similar to that studied in this work shows that variations in soiling rate after mirror washing are difficult to predict [36]. Usually, the reflectivity measurements are locally conducted at a limited number of points on the facets of the collector; these readings are then averaged. However, since the soiling distribution on the collector is often not uniform, it is difficult to assess how representative this average is. In addition, although the reflectivity can be recorded regularly, it is very difficult to carry out continuous measurements during the ongoing operation of the plant. Therefore, the previously described methodology is difficult to implement in the real-time performance monitoring system of a dish-Stirling plant.

Moreover, as described above, the soiling of mirrors not only affects the reflectivity of mirrors but, more generally, reduces the global optical efficiency of the collector. For this reason, in this work, the cleaning index of the mirror $\eta_{cle}$ has been considered, from a conceptual point of view, to be a more correct indicator of the global effect that soiling can induce on the optical performance of the collector compared to the reflectance.

Despite the complexity of the problem, after verifying the great impact of mirror soiling on the performance of the Palermo dish-Stirling (see Fig. 3), we decided to develop a simple and practical numerical method for estimating a daily averaged cleanliness index of mirrors $\eta_{cle}^{day}$. This index has been defined in analogy to what is proposed in the literature to study the soiling level of the modules of photovoltaic systems [46]. The main purpose of the proposed method was to develop a numerical tool for the real-time optimisation of the frequency of washing cycles of the mirrors. The algorithm, described in detail by the calculation flowcharts in Fig. 10, is based on the simple comparison between the measured performance of the system (with soiled mirrors) and that expected using the predictive model previously calibrated using data referring to the condition of clean mirrors of the collector. A similar approach was recently proposed to assess the effect of the soiling of modules on the energy production of large scale photovoltaic plants [45].

Finally, it is important to underline that the method proposed with the algorithm in Fig. 10 should not be intended as an alternative system with respect to traditional *in situ* measurement techniques to estimate the reflectivity of mirrors. Rather, it is a method that can be used continuously to monitor the daily optical efficiency variations and plan reflectometric measurements of mirrors (whenever the calculated index is below a pre-defined threshold) and, if necessary, arrange for the washing of the mirrors.

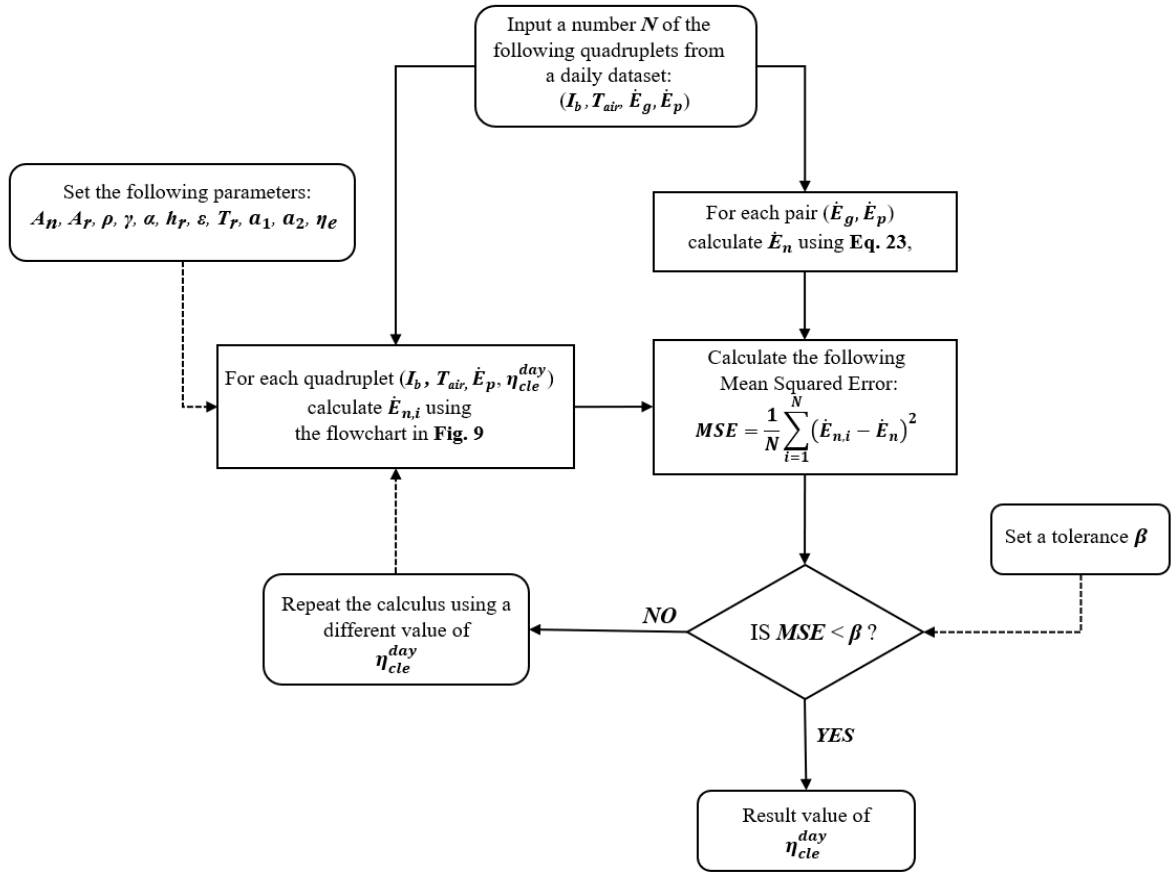

Figure 10. Calculation flowcharts describing the algorithm for the determination of the daily averaged cleanliness index of the solar collector.

## 4. Results and Discussion

### 4.1 *Simulation of the dish-Stirling performance with clean mirrors*

The numerical model described in a compact form by Eq. 26 has been calibrated on the clean mirror data collected for the dish-Stirling collector of Palermo and used to simulate the measured performance of the plant, in order to verify both the predictive capacity of the model and to test the hypotheses underlying it. With this aim, the parameters from Tab. 2 were defined previously using both the technical information made available by *Ripasso Energy* and the results from specific experimental tests carried out on the dish-Stirling unit located in Palermo. The calibration was done by implementing the algorithm described in Fig. 8 using the parameters from Tab. 2 and the measured data from a single day (20 June 2018). On this day, for which the performance has already been represented in Fig. 4, the mirrors of the collector were clean and there was a large variation in *DNI* values (300 - 800 W/m$^2$). Through the calibration procedure, the following parameters from Eq. 19 were estimated: $a_1$ =0.477 and $a_2$ =3.9 kW. Then, using this equation, it was possible to calculate the mechanical power output from the Stirling engine as a function of the thermal power input, estimated through Eq. 12, using both the data from June 20 and that from all the other 12 days when the mirrors of the collector were clean.

Table 2: Parameters estimated for the *Ripasso Energy* dish-Stirling unit.

| Parameters | Value | Units |
|---|---|---|
| Effective aperture area $A_n$ | 101 | m$^2$ |
| Receiver aperture area $A_r$ | 0.0314 | m$^2$ |
| Clean reflectivity of mirrors $\rho$ | 0.95 | - |
| Collector intercept factor $\gamma$ | 0.973 | - |
| Effective receiver absorptance $\alpha$ | 0.92 | - |
| Clean mirrors optical efficiency $\eta_o$ | 0.85 | - |
| Receiver convective coefficient $h_r$ | 10 | W/(m$^2$·K) |
| Receiver effective emissivity $\varepsilon$ | 0.88 | - |
| Alternator electrical efficiency $\eta_e$ | 0.924 | - |

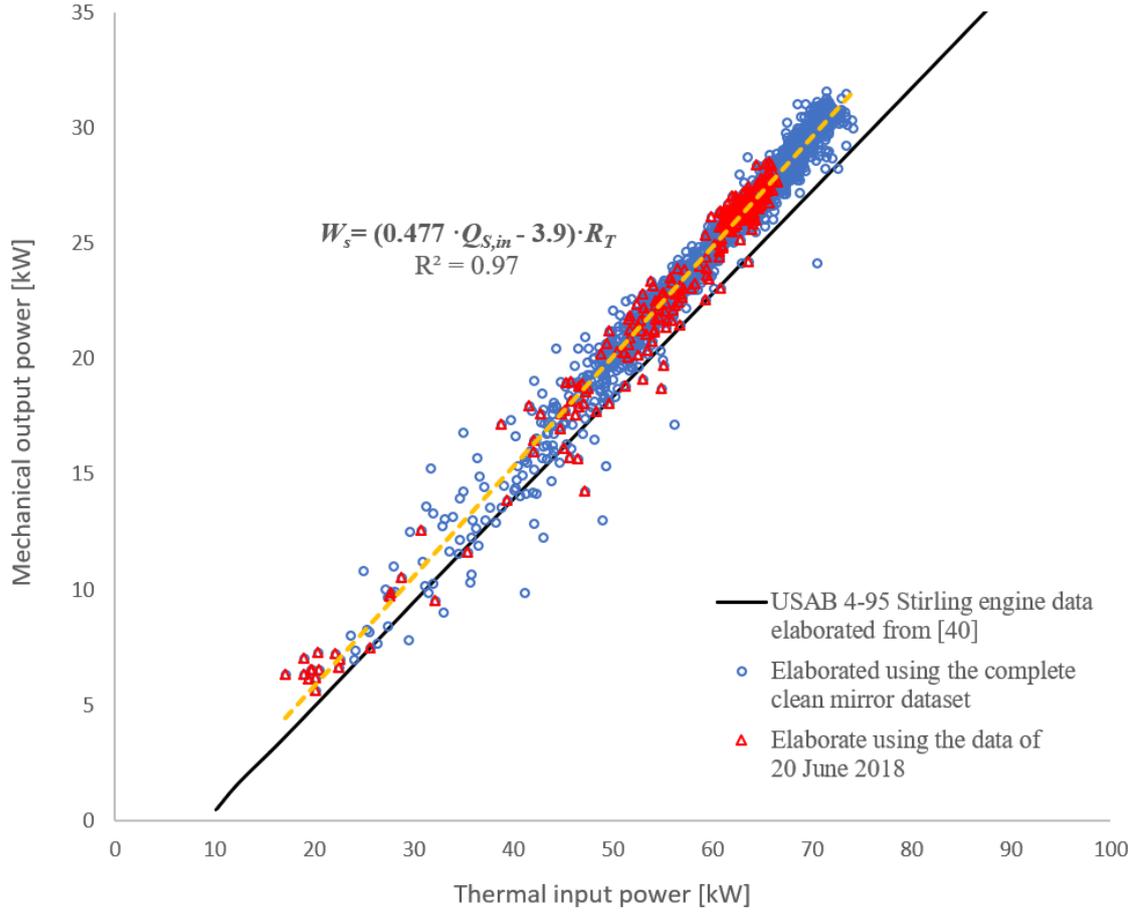

Figure 11. Stirling engine mechanical output power versus thermal input power. The comparison shows that the Ripasso engine represents an optimised version of the original USAB 4-95.

The results of these calculations have been plotted in Fig. 11 alongside the performance data from another Stirling engine, the USAB 4-95, this engine belonging to the same family as the *Ripasso* engine, the latter being an improved version. The data depicted in Fig 11, which is elaborated from [44], refers, in particular, to an advanced model with a rated input power of $\dot{Q}_{S,in}$ =84.2 kW. This engine is equipped with anti-friction bearings and operates with a heater-head temperature of $T_h$=720 °C and 1800 *rpm*.

It is interesting to note, as a result of this first calibration stage, that:

- the values of $\dot{W}_S$ as a function of $\dot{Q}_{S,in}$ calculated using the complete dataset from all 13 days (when the collector mirrors were clean) are strongly correlated by a linear relationship that was calibrated using data from a single day

- real data from Stirling engines operating at partial loads show this similar linear thermal-to-mechanical power relation and the two lines, although independently determined, are close to each other.

These observations reinforce the validity of the hypotheses underlying the approach described in this work.

Taking the results described above, the relationship between the values of $\eta_S$ and $\dot{Q}_{S,in}$, calculated with Eq. 21 using the previously determined values of $\dot{Q}_{S,in}$ and $R_T$ are plotted in Fig. 12 alongside the analogous efficiency curve plot of the *USAB 4-95* engine. The corresponding values of both the Carnot efficiency $\eta_{S,C}$ (calculated by Eq. 17 using the recorded data of $T_h$ and $T_c$) and the exergetic efficiency $\eta_{S,ex}$ (calculated by Eq. 16, Eq. 17 and Eq. 21) are also shown in the same figure as a function of thermal input power $\dot{Q}_{S,in}$. Results from Fig. 12 show that the efficiency of the *Ripasso* Stirling engine $\eta_S$ ranges between 0.28 and 0.43, corresponding with a thermal input power $\dot{Q}_{S,in}$ rising from 17 kW up to 74 kW. On the other hand, corresponding with the same variations of the thermal input power $\dot{Q}_{S,in}$, the Carnot efficiency of the engine $\eta_{S,C}$ ranges between 0.64 and 0.71, with an average value of 0.65.

These values have been calculated considering that during operation, the highest limiting engine temperature was almost constant and equal to $T_h$=993.15 K while the lowest temperature $T_c$ ranged between 293.15 K and 360.15 K. Finally, as depicted in Fig. 12, the resulting variations in the exergetic efficiency of the *Ripasso* engine $\eta_{S,ex}$, calculated as a function of the partialisation of the thermal power delivered to the engine, are between 0.45 and 0.67.

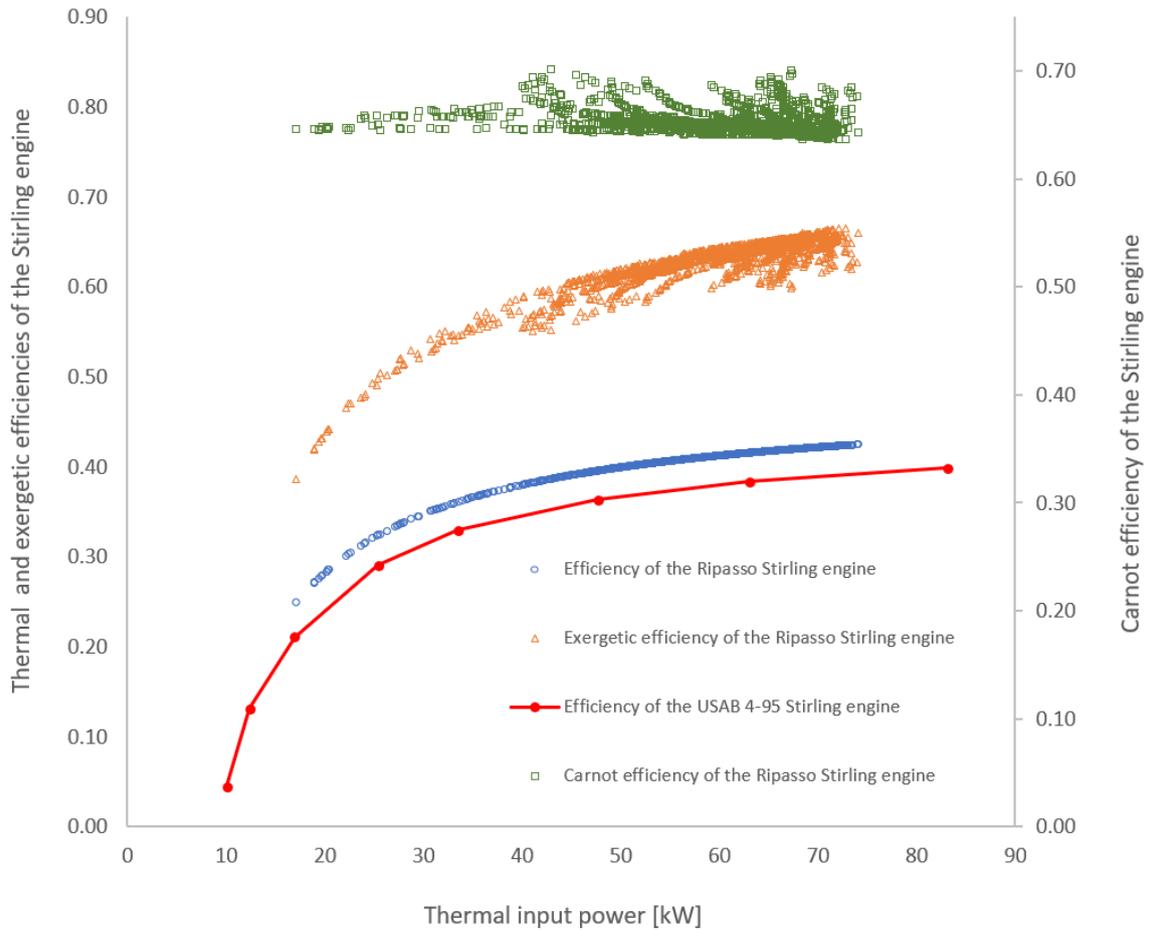

Figure 12. Stirling engine efficiencies versus thermal input power.

The latest results, obtained from real-world operating data of a Stirling engine, are consistent with what is shown in the literature for similar engines [18], [23]. However, it does not seem justified to assume the common hypothesis of considering a constant value for the exergetic efficiency of the engine equal to $\eta_{S,ex}=0.5$ [13].

Once the model was calibrated, it was possible to carry out simulations aimed at predicting the net electricity production of the dish-Stirling plant installed in Palermo. To perform these simulations, the procedure illustrated by the calculation flowchart in Fig. 9 was used. A first test to validate the capabilities of the model was carried out by calculating the rated electrical output power of the plant and comparing this value with that declared by the manufacturer.

Thus, assuming that the collector mirrors are clean, and considering both a solar irradiation of *DNI*=960 W/m² and an external temperature $T_{air}$=25 °C, the model predicts an input power $\dot{Q}_{S,in}$ =80.72 kW, a mechanical output power $\dot{W}_S$ =34.66 kW and a gross electric power $\dot{E}_g$ =32.02 kW. Finally, considering parasitic absorptions only related to the cooling system, approximately equal to $\dot{E}_p$ =1.1 kW, it is possible to calculate a net electric output power equal to $\dot{E}_n$ =30.92 kW. The latter quantity, calculated conforming to the definition of net electric power proposed by *Ripasso Energy*, is very close to the rated electric net power reported in Tab. 1.

Afterwards, the numerical model was used to simulate the net electric output power data recorded during the 13 days when the collector mirrors of the dish-Stirling unit were clean. In order to perform these simulations, the measured values of $T_{air}$ and *DNI* were used. Moreover, the total parasitic absorption power was assumed to equal $\dot{E}_p$ =3.6 kW which is the average of the record measured when the collector mirrors were clean ($\eta_{cle}$ =1). A comparison of measured and predicted values is depicted in Fig. 13, showing the ability of the proposed model to predict the electric power output generated by the plant.

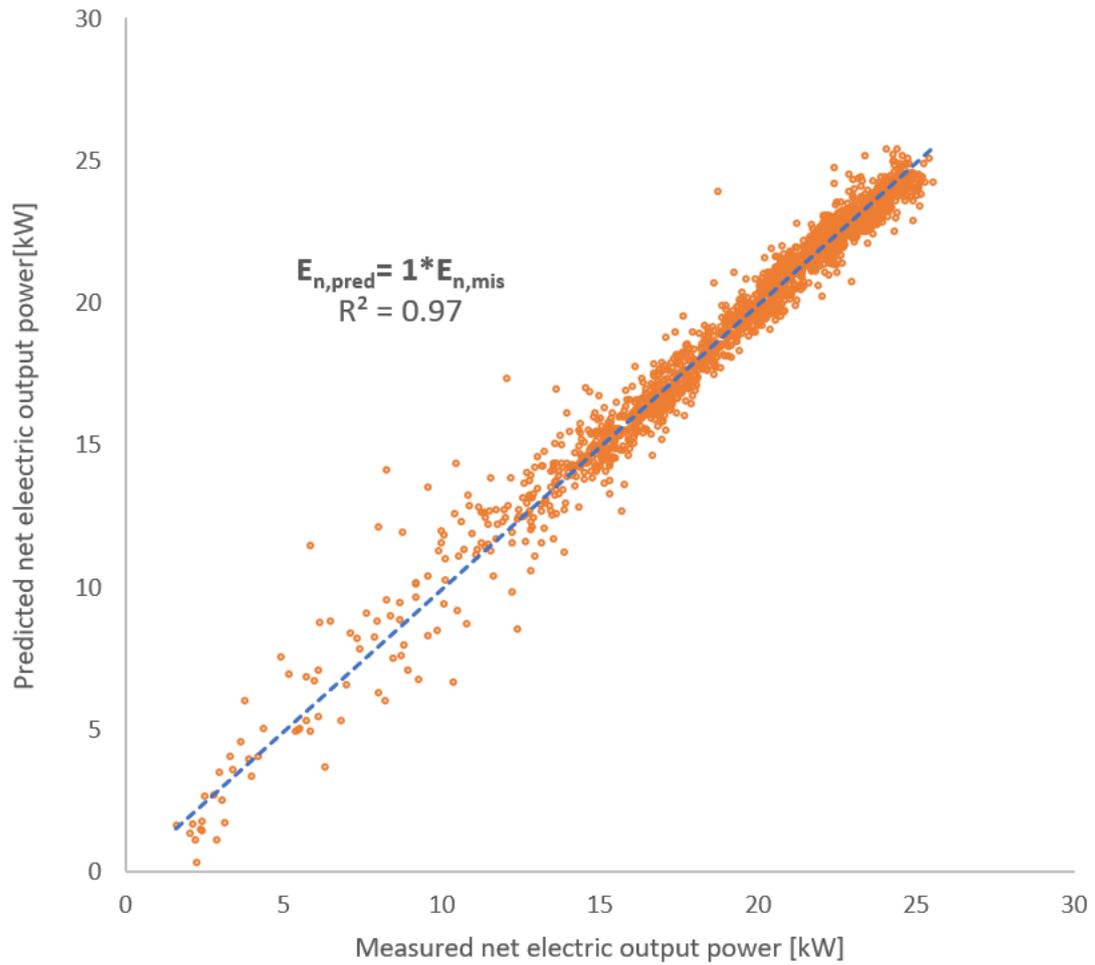

Figure 13. Measured versus predicted net electric output power for the days on which the collector mirrors were clean.

Fig. 14 and 15 show the measured and predicted net electric output power of the system for two clear-sky days, during which the ambient temperature was 15 °C and 25 °C, respectively

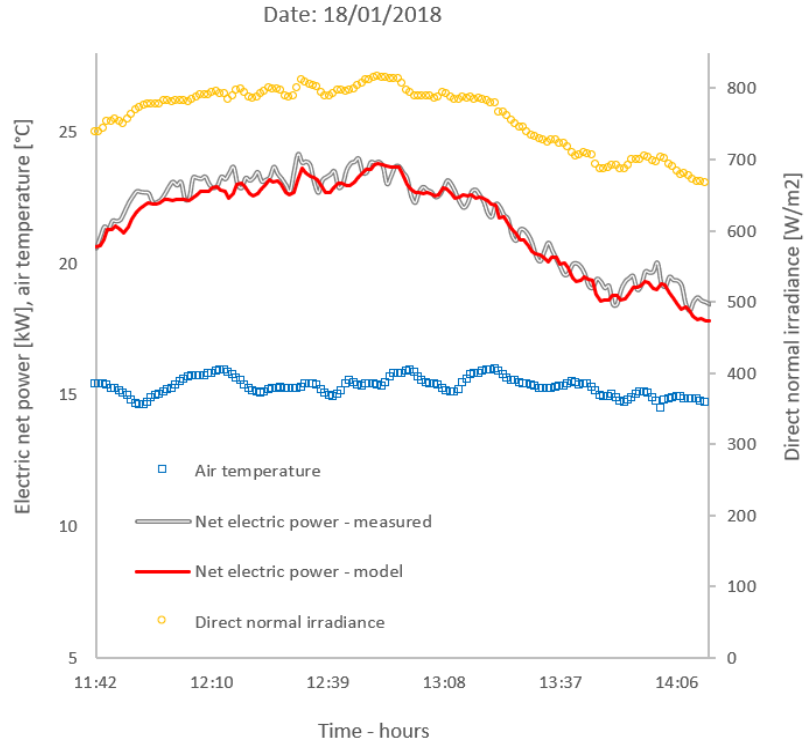

Figure 14. Measured versus predicted net electric power output on 18 January 2018. *DNI* and air temperature variations are also shown.

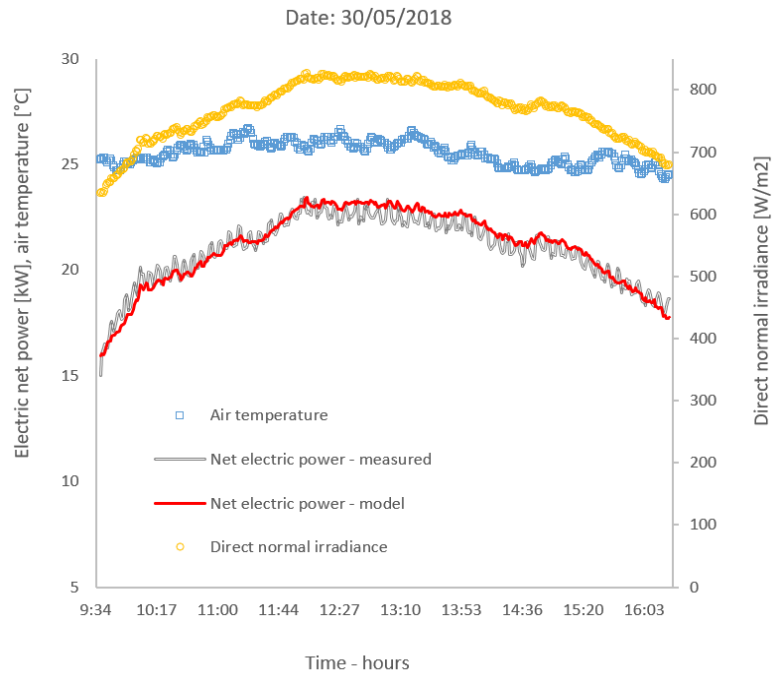

Figure 15. Measured versus predicted net electric power output on 30 May 2018. *DNI* and air temperature variations are also shown.

Finally, using Eqs. 7-10 with the input values of $T_{air}$ from the same dataset, it has been possible to calculate the thermal losses from the collector receiver. These losses present an approximately linear trend with the variations in external temperature, passing from $\dot{Q}_{r,out}=174$ kW at $T_a$=15 °C down to $\dot{Q}_{r,out}=173$ kW at $T_a$ =30 °C as depicted in Fig. 16. Therefore, on the basis of these observations, we noticed that assuming a constant value of the thermal losses from the receiver in Eq. 26, equal to the average of the measurements $\dot{Q}_{r,out}^{ave}$ =173.5 kW, no substantial errors were introduced into the results of the simulations. This important observation allows us to verify the assumption behind the linearised models described by Eqs. 27-29.

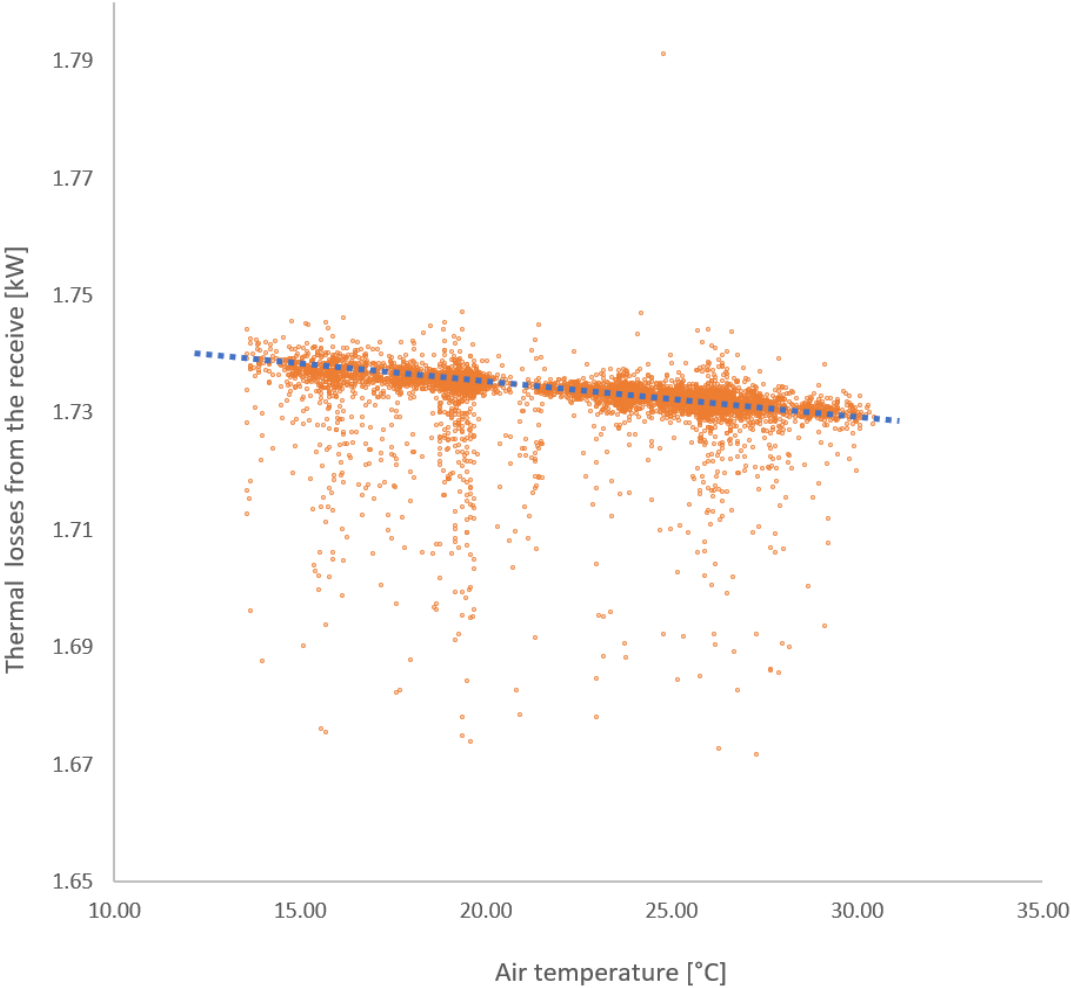

Figure 16. Thermal losses from the receiver versus changes in air temperature recorded during the testing campaign.

4.2 *Example of application of the linearised model.*

To verify the possibilities offered by the simplified linear model described by Eqs. 29-30, we decided to use it to study the performance data of a different plant from that of Palermo: the Stirling Energy System (SES) equipped with the USAB 4-95 Stirling engine [9], [10]. For this system the waterfall chart at peak operation condition ($I_b^{max}$ =1000 W/m$^2$) and the experimental data of net electrical output power as a function of direct normal irradiation, were used. This information, elaborated from the data available in the literature [9], is illustrated by Fig. 17 and Fig. 18, respectively. Thus, both using the waterfall chart in Fig. 17 and fitting the data in Fig. 18 with a linear regression, we have estimated the minimum parameters necessary for subsequent processing These parameters have been reported in Tab. 3. Thus, assuming a value of the receiver absorptivity equal to *α*=0.92 it was possible to estimate $\eta_o$ =0.81 for the system. Finally, from Eqs. 32-34, we obtained *a*$_1$ =0.445 and *a*$_2$ =2.54 kW.

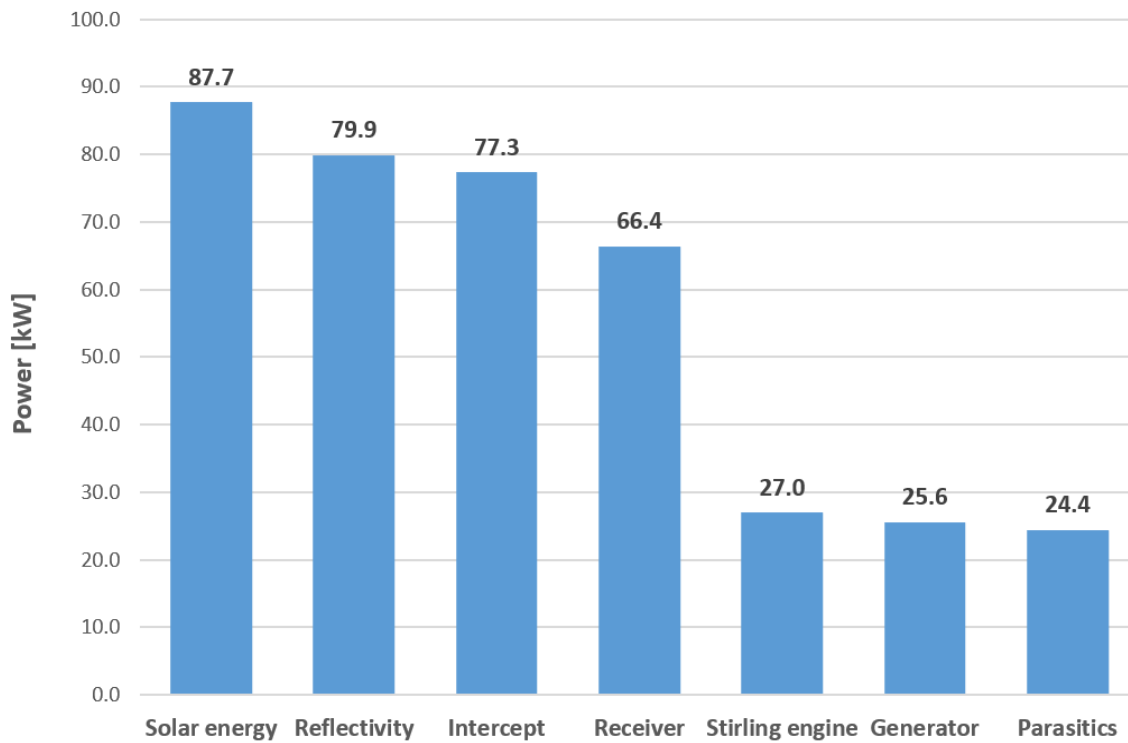

Figure 17. Waterfall chart for SES dish-Stirling (elaborated from [9])

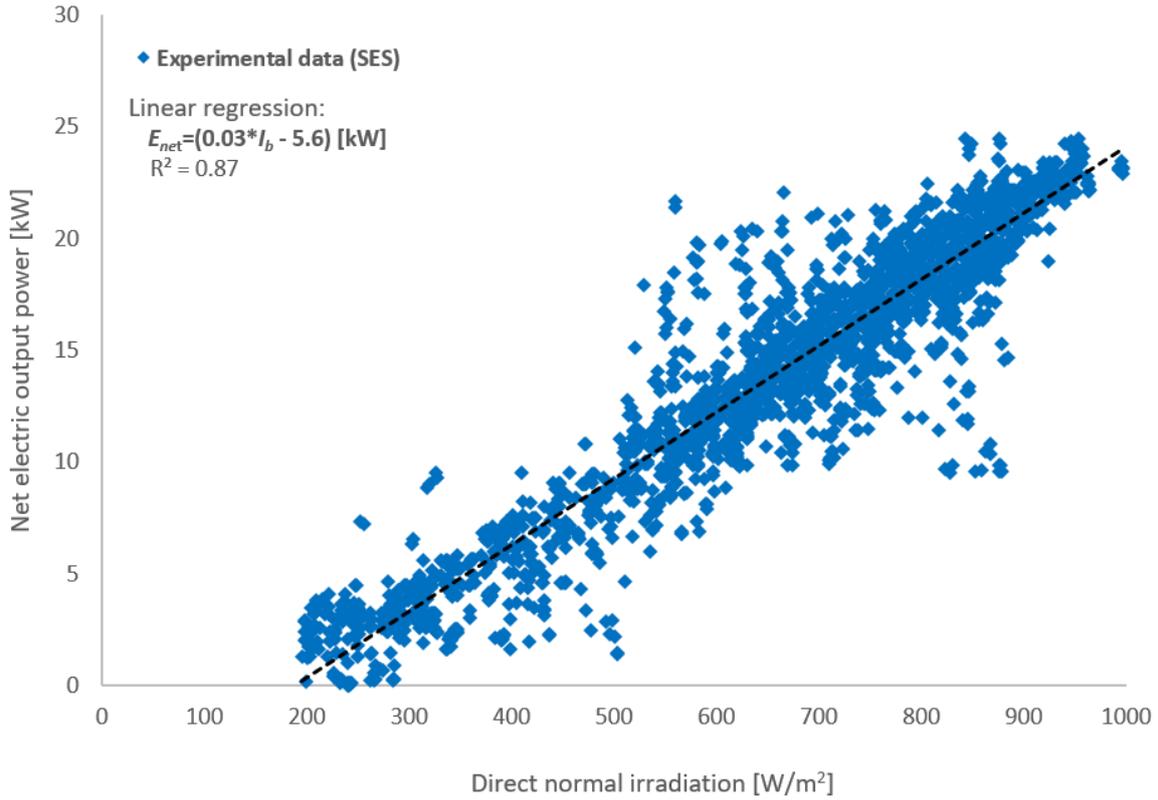

Figure 18. Net output power versus DNI of SES dish-Stirling (elaborated from [9])

Table 3: Parameters estimated for *SES* dish-Stirling unit

| Parameter | Value | Units |
|---|---|---|
| Effective aperture area $A_n$ | 87.7 | m$^2$ |
| Clean reflectivity of mirrors $\rho$ | 0.91 | - |
| Collector intercept factor $\gamma$ | 0.97 | - |
| Alternator electrical efficiency $\eta_e$ | 0.95 | - |
| Maximum design DNI $\dot{I}_b^{max}$ | 1000 | W/m$^2$ |
| Average parasitic absorption $\dot{E}_p^{ave}$ | 1.2 | kW |
| Average receiver thermal losses $\dot{Q}_{r,out}^{ave}$ | 4.72 | kW |
| Slope linear relation $b_1$ | 0.030 | (kW·m$^2$)/W |
| Intercept linear relation $b_2$ | 5.60 | kW |

After the calibration stage it was possible to calculate the relationship between the net electric output power as a function of the variations of the thermal input power to the Stirling engine. These results, along with those elaborated by a previous study on the same system [35], have been plotted in Fig. 19. With regards to this last result, it is interesting to note that there is a good agreement between the two analyses, despite the fact that the present study is based on a simplified approach using the limited number of information available in literature for the SES system.

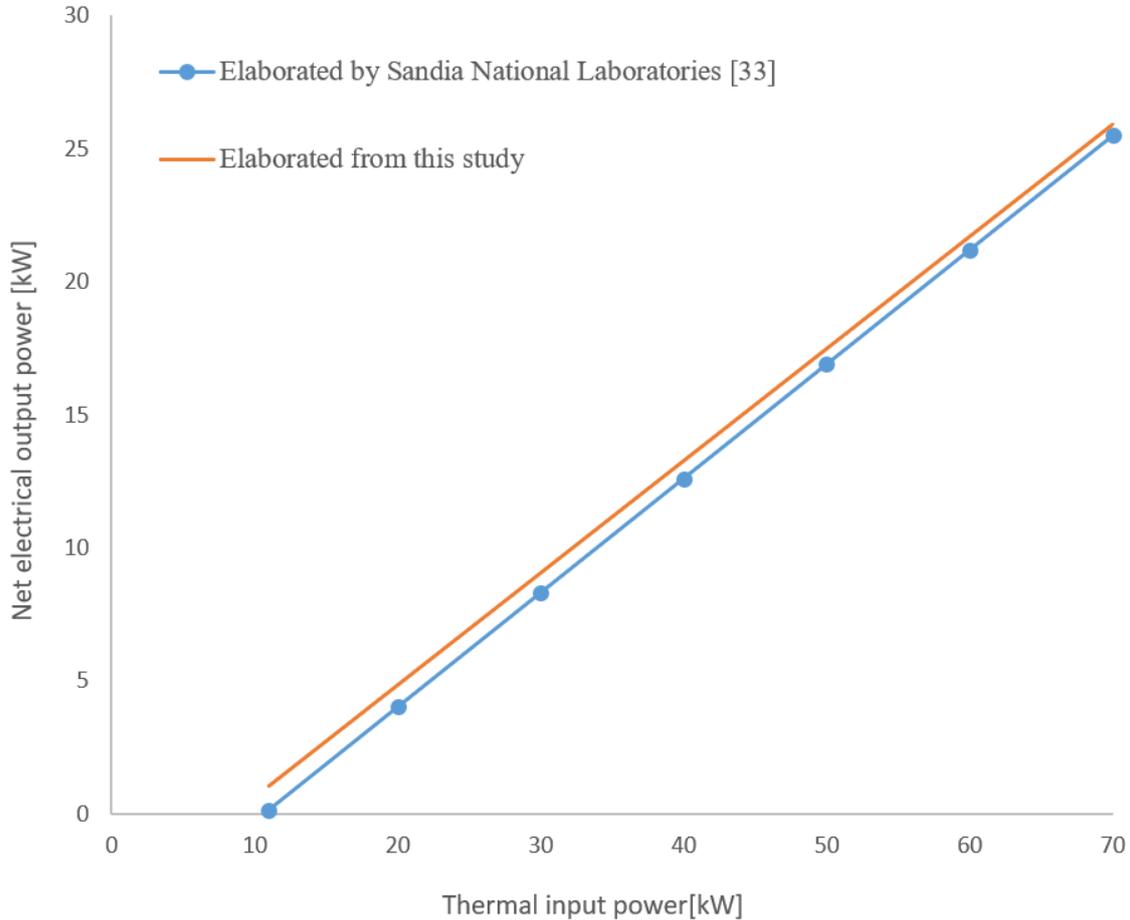

Figure 19. net electric output power versus thermal input power for SES dish-Stirling calculated by the method elaborated in this study and compared with the results elaborated by the Sandia National Laboratories

Finally, we decided to diagram the partial load efficiency curves of both *Ripasso* and SES Stirling engines as they have been extrapolated from this study and to compare them with the experimental curve of the USAB 4-95 engine. These curves have been plotted in Fig. 20 in terms of normalised efficiency versus normalised thermal input power. It is possible to define the normalised efficiency as the ratio of $\eta_S$ over $\eta_S^{max}$ and the normalised thermal input as the ratio of $\dot{Q}_{S,in}$ over $\dot{Q}_S^{max}$. As expected, the normalised efficiency curves of the USAB 4-95, SES and *Ripasso* engines are substantially coincident since they all belong to the same family.

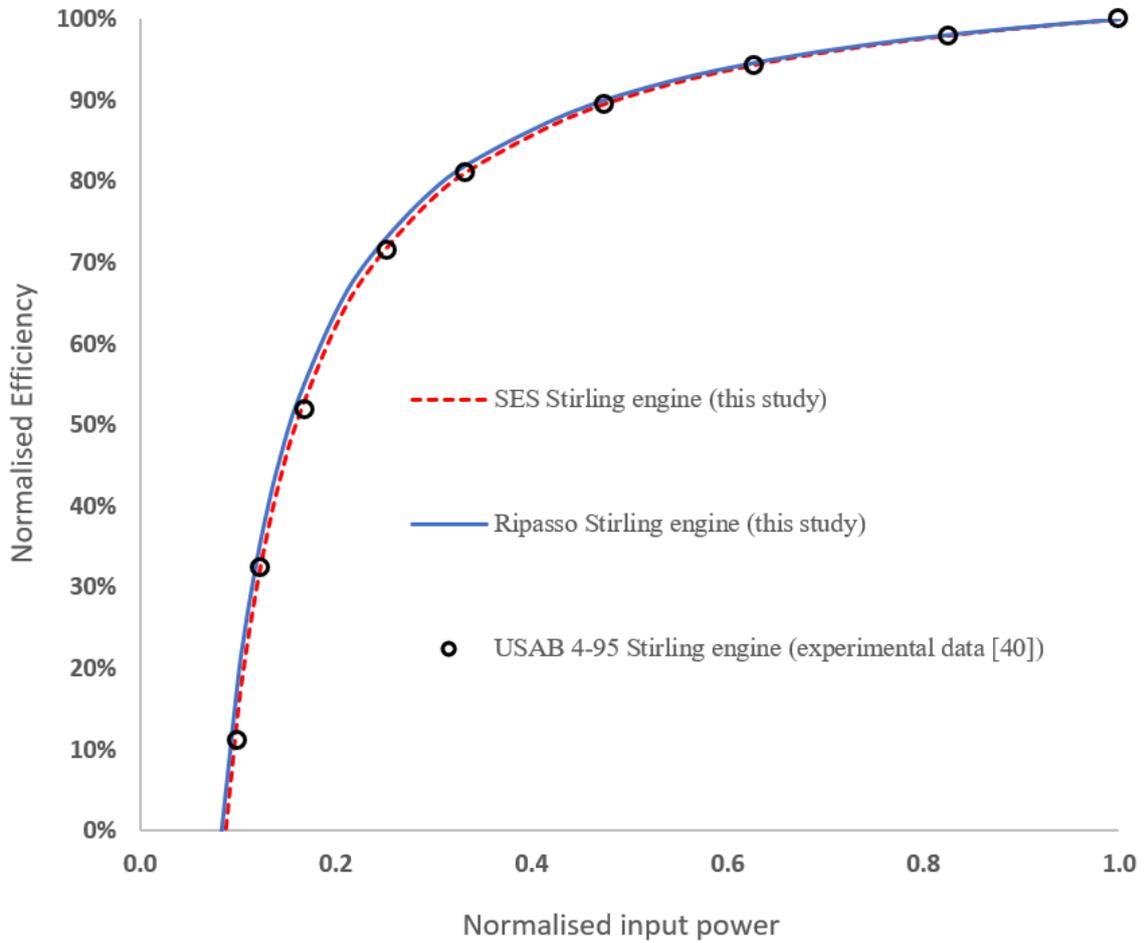

Figure 20. Normalised efficiencies versus normalised input power for the three different Stirling engines (SES, Ripasso and USAB 4-95) in this study.

4.3 *Estimation of the daily averaged cleanliness index of the mirrors*

The results presented so far have mainly concerned the performance of the dish-Stirling plant with clean collector mirrors. The remaining data regarding the periods during which the mirrors were soiled has been analysed using the methodology defined in the calculation flowchart in Fig. 10 with the purpose of defining the time-series plot of the daily average cleanliness index of the mirrors. An example of the application of the proposed method is shown in Fig. 21. In this figure, the net electric power curve predicted by the model assuming clean mirrors is plotted along with the curve corresponding to the condition of soiled mirrors. The application of the algorithm described in Fig. 10 has made it possible to estimate, for this particular day, a value for the daily average cleanliness index equal to $\eta_{cle}^{day} = 0.85$.

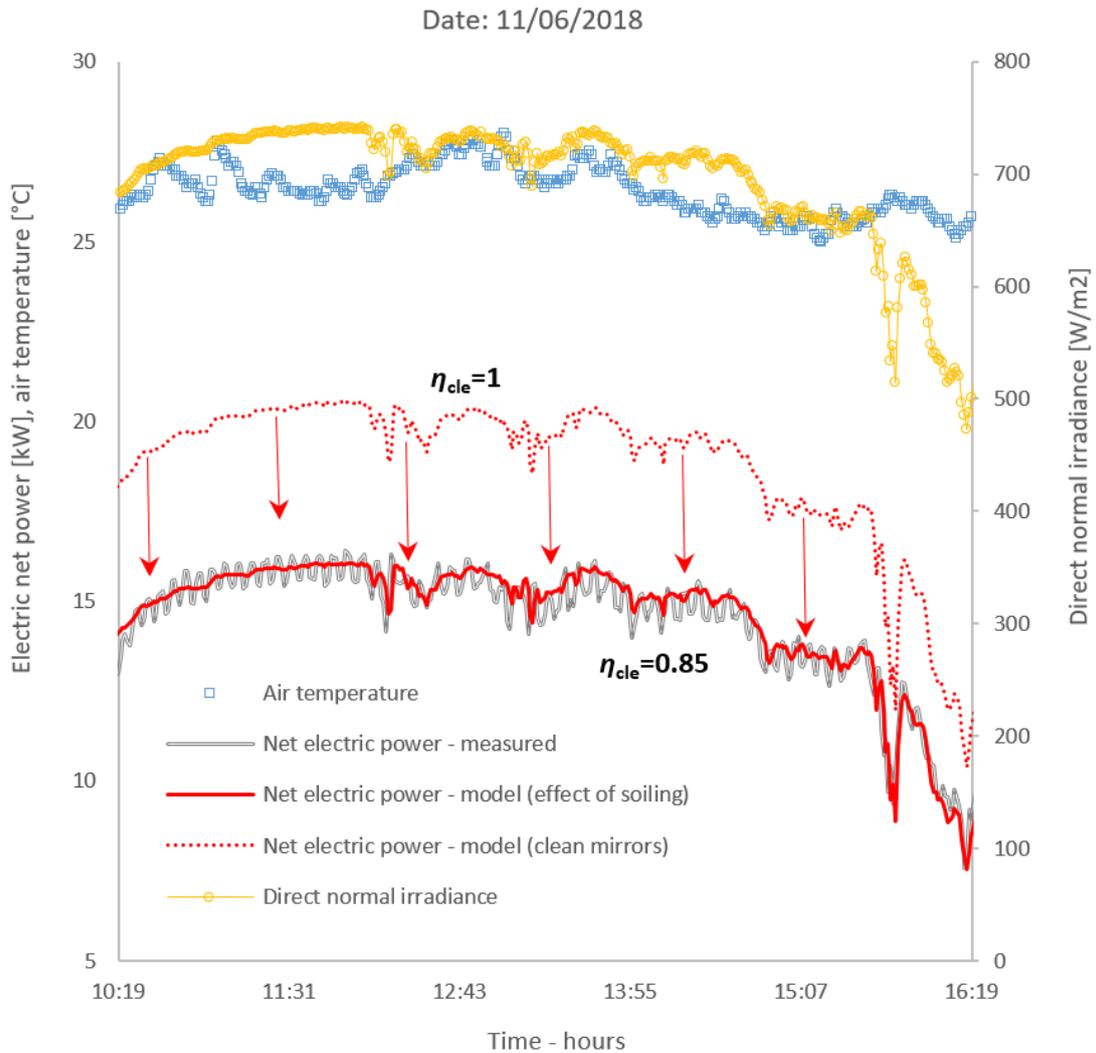

Figure 21. Determination of the daily average cleanliness index of mirrors on 11 June 2018 using the algorithm outlined in Fig. 10.

The results of all the elaborations are shown in Fig. 22. This figure shows the variation in the calculated cleanliness index function against the rate of rainfall, expressed as (mm/day), and significant Saharan dust transport events (Sirocco), which are highlighted with yellow bars. The pattern of the calculated variations is similar to measurements of reflectivity history for a dish-Stirling system described in the literature [36]. It is possible to observe in Fig. 22 that during the first few months of monitoring, an intense period of rainfall contributed to the maintenance of clean mirrors; however, from point *A* to point *B*, there was a notable drop in the cleanliness index to 0.73, due to a prolonged Sirocco event from 28 February until 5 March (see Fig. 23). Immediately afterwards, a period of rainfall between 7 and 12 March corresponding to points *B* and *C* on the graph, was echoed by a rise in the cleanliness coefficient to 0.96. From points *C* to *D*, there was a gradual reduction in the cleanliness index and then a subsequent increase due to other rain events (point *E*). In the period between the points *E* and *F* on the graph, there is no data since the plant was off. However, it is plausible that during this period a pattern similar to that between *A* and *C* occurred as a consequence of a second Sirocco event from 21 to 24 March (see Fig. 24) and a subsequent period of rain. Gradual soiling continued from point *F* to *G* since during this time there were no significant weather events. The next Sirocco event, which was much shorter in duration than the first, was recorded in April from 15 to 17 (see Fig. 25). During this event, the cleanliness index dropped to 0.82, point *H*, and after the subsequent rains rose again to 0.93, point *I*. Approximately one month passed before manual cleaning and maintenance of the dish was performed, point *J*. During this time, the rate of soiling was slow and steady, and indeed no other significant weather events were recorded. A marked difference in the cleanliness index can be seen after the mechanical cleaning of 23 May, with this number rising from 0.82 to 1.0 (point *K*). There was a final cycle of soiling and washing of the mirrors since a soil excavation, that was going on near the plant (between 31 March and 8 June) produced the last notable drop in the cleanliness index and a subsequent sequence of heavy rains (occurring on 14-16 June) produced a complete washing of the collector mirrors (points *L* to *M*). The soil excavation period is highlighted with grey bars on the graph.

In conclusion, the application of the proposed method has made it possible to obtain a time-series plot of the daily average cleanliness index, which is congruent with the repeated cycles of rainfall and Saharan dust transport events. Moreover, these results confirm what is already known in the literature: wet deposition is the dominant mechanism and is responsible for about 80% of the deposited dust. However, wet deposition is concentrated on rainy days which are quite few per year (such as between points *A* and *B* in Fig. 22). Dry deposition, on the other hand, is less dominant but it occurs almost every day [54] (such as between points *I* and *J* in Fig. 22).

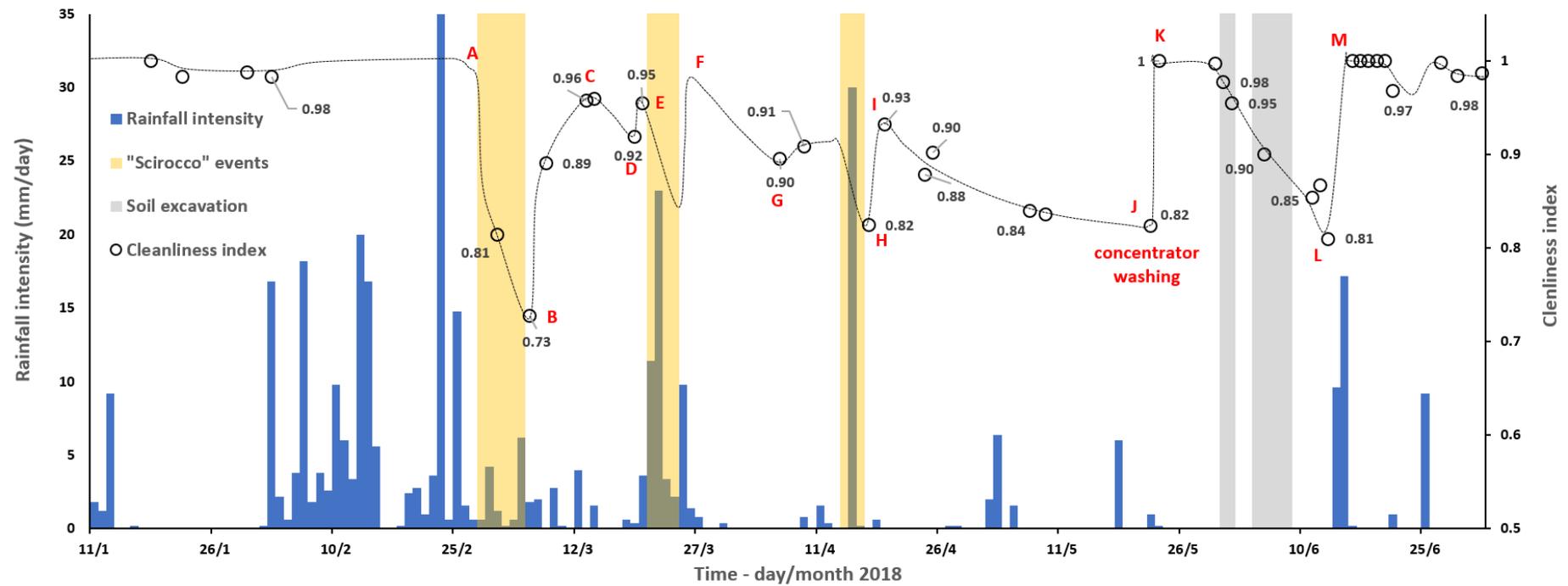

Figure 22. Time-series plot of the daily average cleanliness index, rainfall and sirocco events.

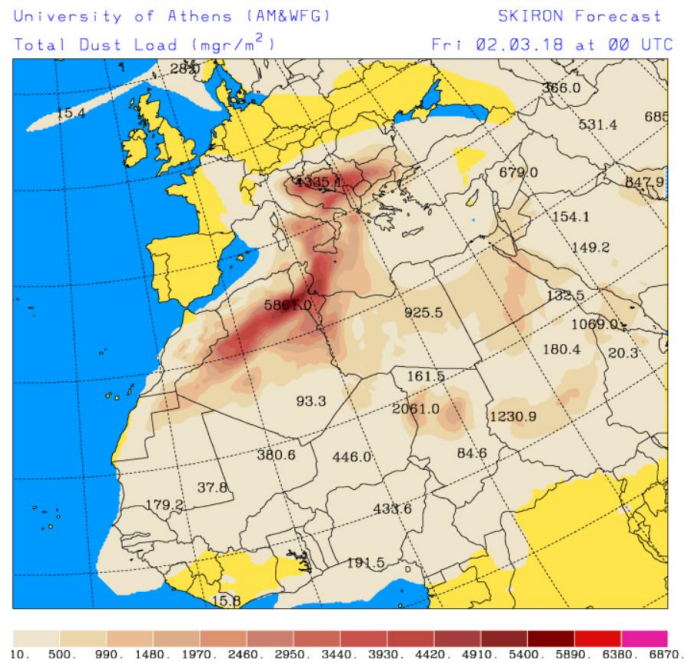

Figure 23. Total dust load (mrg/m$^2$) predicted with SKIRON/Dust system on 2 March 2018

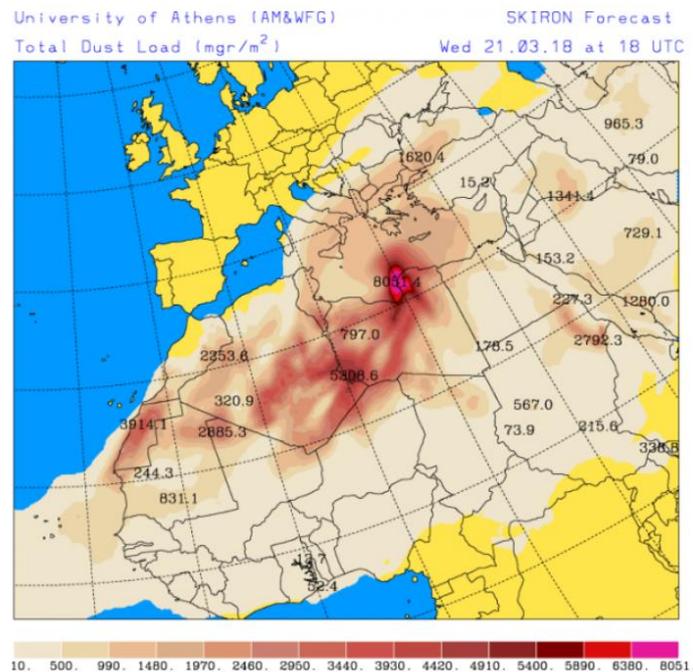

Figure 24. Total dust load (mrg/m$^2$) predicted with SKIRON/Dust system on 21 March 2018

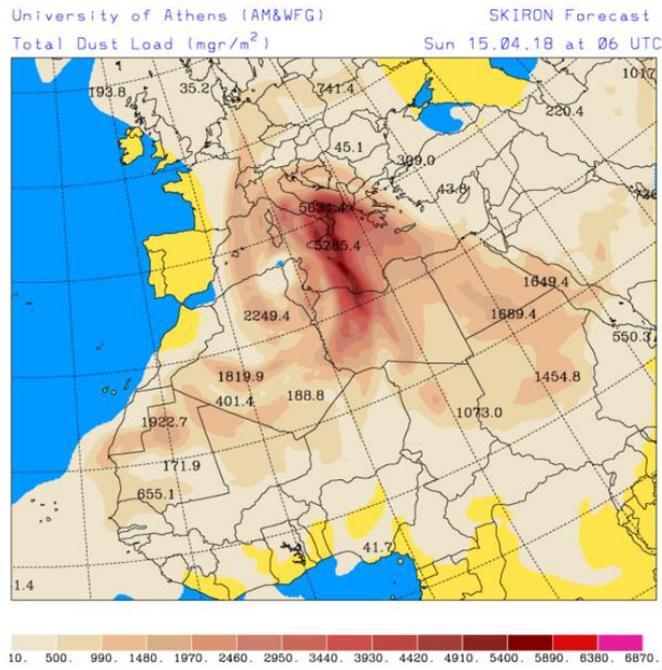

Figure 25. Total dust load (mrg/m$^2$) predicted with SKIRON/Dust system on 15 April 2018

With the aim of having a summary picture of all the elaborations, all the experimental data of output power versus *DNI* were aggregated in three intervals of cleanliness index and plotted in Fig. 26, along with the lines representing the output of the model at fixed indices.

The research activities planned for the plant in the future will provide for the automation of the proposed method, which will be integrated with a campaign of periodic measurements of mirror reflectivity in order to consolidate this approach, which will be ultimately aimed at the economic optimisation of the activities of cyclical washing of the collector mirrors.

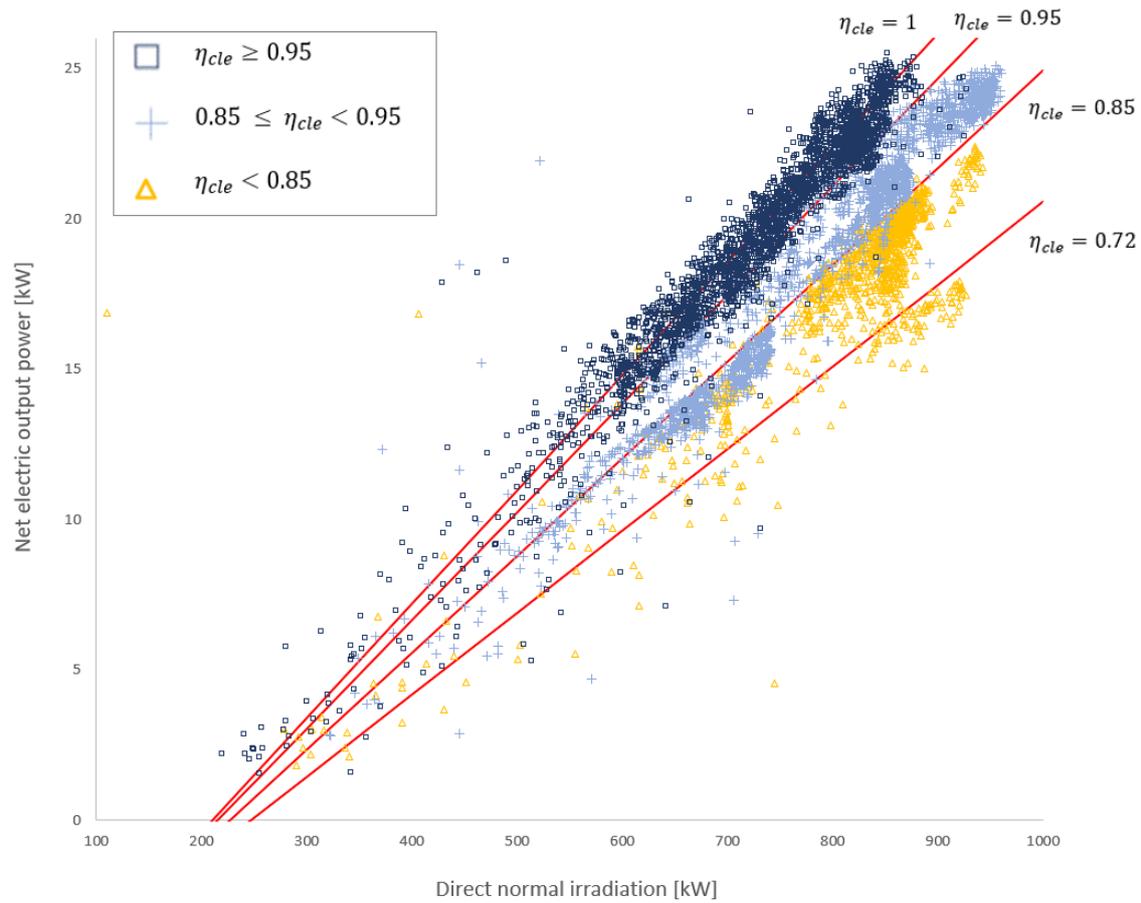

Figure 26. Net electric power output versus DNI (experimental data aggregated in three different cleaning index intervals and model lines at different fixed values of the same index)

## 5. Conclusion

In this paper, we presented a numerical model for the analysis of energy production of dish-Stirling systems that was validated with experimental data from a 32 kW$_e$ solar plant recently installed at a facility test site on the University of Palermo campus, in the south of Italy. The model was developed by introducing a simplified expression of the Stirling engine part load efficiency into the set of equations describing the energy balance of the system and was calibrated using data from a single day of plant operation of the Palermo dish-Stirling plant. The results of the simulations carried out with the model show excellent agreement with the experimental data of energy production corresponding to the condition of clean collector mirrors.

Moreover, the model can be linearised by introducing further simplifying assumptions based on the experimental evidence. The resulting analytical expression, which is a linear relationship between solar irradiation and electric power output from the system, for the first time suggests a theoretical interpretation of what is extensively observed in literature data and which is the basis for the empirical model proposed by Stine for dish-Stirling plants. The main advantage to using the proposed model in this work compared to similar empirical expression is that it is possible to perform optimisation analyses of dish-Stirling components taking into account the main parameters that govern the overall efficiency of the system, including the operating curve of the Stirling engine.

The model was also used to define a numerical procedure for assessing the effect of mirror soiling on the optical efficiency of the solar collector that can be integrated with periodic measurements of mirror reflectivity in order to optimise of the activities of cyclical washing of the collector mirrors. The application of the proposed method to the data collected during the testing campaign of the solar collector of Palermo, has made it possible to obtain a time-series plot of the daily average cleanliness index, which is congruent with the repeated cycles of rainfall and Saharan dust transport events.

In future works, the numerical approach proposed in this paper will be used to assess the monthly and annual energy performance of dish-Stirling plants built in the central Mediterranean area, taking into account the particular climatic conditions of this geographical location. Such studies are of fundamental importance for the development of guidelines that can direct national policies in favouring a greater commercial penetration of this interesting type of renewable energy production system.

**Acknowledgements**

The authors express gratitude to the companies *HorizonFirm S.r.l.*, *Elettrocostruzioni S.r.l.* and *Ripasso Energy* for the support provided without which it would not have been possible to install the dish-Stirling concentrator plant at the University of Palermo campus.

**Nomenclature**

$A_a$     total collector aperture area [m$^2$]

| Symbol | Description |
|---|---|
| $A_n$ | effective reflective surface of the collector (projected mirror area) [m$^2$] |
| $A_r$ | receiver equivalent surface area [m$^2$] |
| $a_1$ | slope of the linear relation between $\dot{W}_S$ and $\dot{Q}_{S,in}$ (dimensionless) |
| $a_2$ | intercept of the linear the relationship between $\dot{W}_S$ and $\dot{Q}_{S,in}$ [W] |
| $b_1$ | slope of the linear relationship between $\dot{E}_n$ and $I_b$ [kW·m$^2$/W] |
| $b_2$ | intercept of the linear relationship between $\dot{E}_n$ and $I_b$ [W] |
| $C_g$ | geometric concentration ratio of the collector (dimensionless) |
| $\dot{E}_g$ | gross electric output power [W] |
| $\dot{E}_n$ | net electric output power [W] |
| $\dot{E}_p$ | total parasitic absorption power [W] |
| $\dot{E}_{p,d}$ | parasitic absorption power of the dry-cooler and circulation pumps [W] |
| $\dot{E}_{p,t}$ | parasitic absorption power of the tracking system [W] |
| $f_s$ | sampling frequency [Hz] |
| $f_c$ | cut-off frequency [Hz] |
| $h_r$ | natural convective coefficient at receiver surface [W/(m$^2$·K)] |
| $I_b$ | solar beam radiation [W/m$^2$] |
| $\dot{Q}_{con}$ | rate of thermal convective losses from the receiver [W] |
| $\dot{Q}_{rad}$ | flux of radiant energy emitted from the receiver [W] |
| $\dot{Q}_{r,in}$ | thermal energy power absorbed by the receiver [W] |
| $\dot{Q}_{r,out}$ | thermal losses from the receiver [W] |

| Symbol | Description |
|---|---|
| $\dot{Q}_{sun}$ | rate of solar energy incident on the mirrors of the collector [W] |
| $\dot{Q}_{S,in}$ | thermal power delivered to the Stirling engine [W] |
| $\dot{Q}_{S,out}$ | thermal power rejected from the Stirling engine [W] |
| $R_T$ | ratio of temperature $T_o$ to temperature $T_{air}$ (dimensionless) |
| $\dot{W}_S$ | rate of mechanical work of the Stirling engine [W] |
| $T_{air}$ | air temperature [K] |
| $T_o$ | reference temperature [K] |
| $T_c$ | rejection temperature of the engine working fluid [K] |
| $T_r$ | temperature of the receiver surface [K] |
| $T_h$ | heat input temperature of the engine working fluid [K] |
| $T_{sky}$ | effective sky temperature [K] |
| $\eta_c$ | thermal efficiency of the solar collector (dimensionless) |
| $\eta_e$ | average electric generator efficiency (dimensionless) |
| $\eta_{cle}$ | cleanliness index of the collector mirror (dimensionless) |
| $\eta_m$ | factor accounting for the losses of reflective mirror area (dimensionless) |
| $\eta_o$ | effective optical efficiency (dimensionless) |
| $\eta_S$ | efficiency of the Stirling engine (dimensionless) |
| $\eta_{S,C}$ | Carnot cycle efficiency of the Stirling engine (dimensionless) |
| $\eta_{S,ex}$ | exergetic efficiency of the Stirling engine (dimensionless) |
| $N$ | number of samples in the data record (dimensionless) |

*Superscripts*

*ave*  averaged value

*day*  mean daily value

*max*  maximum design value

*Greek letters*

$\alpha$  effective absorptivity of the receiver (dimensionless)

$\varepsilon$  effective emissivity factor of the receiver (dimensionless)

$\rho$  clean mirror reflectance (dimensionless)

$\gamma$  intercept factor of the solar collector (dimensionless)

$\sigma$  Stefan–Boltzmann constant, $5.67\times10^{-8}$ [W/(m$^2$·K$^4$)]

**References**

[1]  Bilgili M, Ozbek A, Sahin B, Kahraman A. An overview of renewable electric power capacity and progress in new technologies in the world. Renew Sustain Energy Rev 2015;49:323–34. doi:10.1016/j.rser.2015.04.148.

[2]  Joint Research Centre. Energy Technology Reference Indicator projections for 2010-2050. 2014. doi:10.2790/057687.

[3]  Zhang HL, Baeyens J, Degrève J, Cacères G. Concentrated solar power plants: Review and design methodology. Renew Sustain Energy Rev 2013;22:466–81. doi:10.1016/j.rser.2013.01.032.

[4]  Zhu S, Yu G, Ma Y, Cheng Y, Wang Y, Yu S, et al. A free-piston Stirling generator integrated with a parabolic trough collector for thermal-to-electric conversion of solar energy. Appl Energy 2019;242:1248–58. doi:10.1016/j.apenergy.2019.03.169.

[5]  Buscemi A, Panno D, Ciulla G, Beccali M, Lo Brano V. Concrete thermal energy storage for linear Fresnel collectors: Exploiting the South Mediterranean's solar potential for agri-food processes. Energy Convers Manag 2018;166. doi:10.1016/j.enconman.2018.04.075.

[6]  Ogunmodimu O, Okoroigwe EC. Concentrating solar power technologies for solar thermal grid electricity in Nigeria: A review. Renew Sustain Energy Rev

2018;90:104–19.

[7]     Sharma A, Shukla SK, Rai AK. Finite time thermodynamic analysis and optimization of solar-dish Stirling heat engine with regenerative losses. Therm Sci 2011;15:995–1009. doi:10.2298/TSCI110418101S.

[8]     Thombare DG, Verma SK. Technological development in the Stirling cycle engines. Renew Sustain Energy Rev 2008;12:1–38. doi:10.1016/j.rser.2006.07.001.

[9]     Mancini T, Heller P, Butler B, Osborn B, Schiel W, Goldberg V, et al. Dish-Stirling Systems: An Overview of Development and Status. J Sol Energy Eng 2003;125:135. doi:10.1115/1.1562634.

[10]    Coventry J, Andraka C. Dish systems for CSP. Sol Energy 2017;152:140–70. doi:10.1016/J.SOLENER.2017.02.056.

[11]    Ferreira AC, Nunes ML, Teixeira JCF, Martins LASB, Teixeira SFCF. Thermodynamic and economic optimization of a solar-powered Stirling engine for micro-cogeneration purposes. Energy 2016;111:1–17. doi:10.1016/j.energy.2016.05.091.

[12]    Reader GT, Hooper C. Stirling Engines. 1983. London E FN Spon n.d.

[13]    Al-Dafaie AMA, Dahdolan ME, Al-Nimr MA. Utilizing the heat rejected from a solar dish Stirling engine in potable water production. Sol Energy 2016;136:317–26. doi:10.1016/j.solener.2016.07.007.

[14]    Hafez AZ, Soliman A, El-Metwally KA, Ismail IM. Design analysis factors and specifications of solar dish technologies for different systems and applications. Renew Sustain Energy Rev 2017. doi:10.1016/j.rser.2016.09.077.

[15]    Singh UR, Kumar A. Review on solar Stirling engine: Development and performance. Therm Sci Eng Prog 2018. doi:10.1016/j.tsep.2018.08.016.

[16]    Bădescu V. Note concerning the maximal efficiency and the optimal operating temperature of solar converters with or without concentration. Renew Energy 1991;1:131–5. doi:10.1016/0960-1481(91)90114-5.

[17]    Chen L, Sun F, Wu C. Optimum collector temperature for solar heat engines. Int J Ambient Energy 1996;17:73–8. doi:10.1080/01430750.1996.9675221.

[18]    Kongtragool B, Wongwises S. Optimum absorber temperature of a once-reflecting full conical concentrator of a low temperature differential Stirling engine. Renew Energy 2005;30:1671–87. doi:10.1016/j.renene.2005.01.003.

[19]    Ahmadi MH. Investigation of Solar Collector Design Parameters Effect onto Solar

Stirling Engine Efficiency. J Appl Mech Eng 2012;01:10–3. doi:10.4172/2168-9873.1000102.

[20] Reddy VS, Kaushik SC, Tyagi SK. Exergetic analysis and performance evaluation of parabolic trough concentrating solar thermal power plant (PTCSTPP). Energy 2012. doi:10.1016/j.energy.2012.01.023.

[21] Duffie JA, Beckman WA. Solar engineering of thermal processes. John Wiley & Sons; 2013.

[22] Ahmadi MH, Sayyaadi H, Mohammadi AH, Barranco-Jimenez MA. Thermo-economic multi-objective optimization of solar dish-Stirling engine by implementing evolutionary algorithm. Energy Convers Manag 2013;73:370–80. doi:10.1016/j.enconman.2013.05.031.

[23] Carrillo Caballero GE, Mendoza LS, Martinez AM, Silva EE, Melian VR, Venturini OJ, et al. Optimization of a Dish Stirling system working with DIR-type receiver using multi-objective techniques. Appl Energy 2017. doi:10.1016/j.apenergy.2017.07.053.

[24] Beltrán-Chacon R, Leal-Chavez D, Sauceda D, Pellegrini-Cervantes M, Borunda M. Design and analysis of a dead volume control for a solar Stirling engine with induction generator. Energy 2015. doi:10.1016/j.energy.2015.09.046.

[25] Hafez AZ, Soliman A, El-Metwally KA, Ismail IM. Solar parabolic dish Stirling engine system design, simulation, and thermal analysis. Energy Convers Manag 2016. doi:10.1016/j.enconman.2016.07.067.

[26] Mendoza Castellanos LS, Carrillo Caballero GE, Melian Cobas VR, Silva Lora EE, Martinez Reyes AM. Mathematical modeling of the geometrical sizing and thermal performance of a Dish/Stirling system for power generation. Renew Energy 2017. doi:10.1016/j.renene.2017.01.020.

[27] Power CS. Technology Roadmap Concentrating Solar Power. Current 2010. doi:10.1787/9789264088139-en.

[28] Guo S, Liu Q, Sun J, Jin H. A review on the utilization of hybrid renewable energy. Renew Sustain Energy Rev 2018. doi:10.1016/j.rser.2018.04.105.

[29] Bravo Y, Carvalho M, Serra LM, Monné C, Alonso S, Moreno F, et al. Environmental evaluation of dish-Stirling technology for power generation. Sol Energy 2012;86:2811–25. doi:10.1016/j.solener.2012.06.019.

[30] Schiel W, Schweiber A, Stine WB. Evaluation of the 9-kw e dish/stirling system of schlaich bergermann und partner using the proposed iea dish/stirling performance analysis guidelines. Intersoc. Energy Convers. Eng. Conf. 1994, 1994.

[31] Stine WB. Experimentally Validated Long-Term Energy Production Prediction Model for Solar Dish/Stirling Electric Generating Systems. In: D.Y. Goswami, L.D. Kannberg, T.R. Mancini, S. Somasundaram eds., editor. Proc. Intersoc. Energy Convers. Eng. Conf. vol. 2, New York, NY, USA: American Society of Mechanical Engineers; 1995, p. 491–5. doi:AC04-94AL85000.

[32] Igo J, Andraka CE. Solar dish field system model for spacing optimization. Proc. Energy Sustain. Conf. 2007, 2007. doi:10.1115/ES2007-36154.

[33] Nepveu F, Ferriere A, Bataille F. Thermal model of a dish/Stirling systems. Sol Energy 2009. doi:10.1016/j.solener.2008.07.008.

[34] Mendoza Castellanos LS, Galindo Noguera AL, Carrillo Caballero GE, De Souza AL, Melian Cobas VR, Silva Lora EE, et al. Experimental analysis and numerical validation of the solar Dish/Stirling system connected to the electric grid. Renew Energy 2019. doi:10.1016/j.renene.2018.11.095.

[35] Andraka CE. Cost/performance tradeoffs for reflectors used in solar concentrating dish systems. 2008 Proc. 2nd Int. Conf. Energy Sustain. ES 2008, 2009.

[36] Lopez CW, Stone KW. Performance of the Southern California Edison Company Stirling Dish. NASA STI/Recon Tech Rep N 1993;94.

[37] Panno D, Buscemi A, Beccali M, Chiaruzzi C, Cipriani G, Ciulla G, et al. A solar assisted seasonal borehole thermal energy system for a non-residential building in the Mediterranean area. Sol Energy 2018. doi:10.1016/j.solener.2018.06.014.

[38] Sciortino L, Agnello S, Barbera M, Bonsignore G, Buscemi A, Candia R, et al. Direct sunlight facility for testing and research in HCPV. AIP Conf. Proc., vol. 1616, 2014, p. 158–61.

[39] Kallos G, Astitha M, Katsafados P, Spyrou C. Long-Range Transport of Anthropogenically and Naturally Produced Particulate Matter in the Mediterranean and North Atlantic: Current State of Knowledge. J Appl Meteorol Climatol 2007;46:1230–51. doi:10.1175/JAM2530.1.

[40] Kallos G, Spyrou C, Astitha M, Mitsakou C, Solomos S, Kushta J, et al. Ten-year operational dust forecasting – Recent model development and future plans. IOP Conf. Ser. Earth Environ. Sci., vol. 7, 2009, p. 12012. doi:10.1088/1755-1307/7/1/012012.

[41] Kallos G, Solomos S, Kushta J, Mitsakou C, Spyrou C, Bartsotas N, et al. Natural and anthropogenic aerosols in the Eastern Mediterranean and Middle East: Possible impacts. Sci Total Environ 2014;488–489:389–97. doi:10.1016/j.scitotenv.2014.02.035.

[42] Spyrou C, Kallos G, Mitsakou C, Athanasiadis P, Kalogeri C. The Effects of Naturally Produced Dust Particles on Radiative Transfer. Adv. Meteorol. Climatol. Atmos. Phys., Springer; 2013, p. 317–23.

[43] Hachicha AA, Yousef BAA, Said Z, Rodríguez I. A review study on the modeling of high-temperature solar thermal collector systems. Renew Sustain Energy Rev 2019. doi:10.1016/j.rser.2019.05.056.

[44] Selcuk MK, Fujita T. A nomographic methodology for use in performance trade-off studies of parabolic dish solar power modules 1984.

[45] Massi Pavan A, Mellit A, De Pieri D. The effect of soiling on energy production for large-scale photovoltaic plants. Sol Energy 2011. doi:10.1016/j.solener.2011.03.006.

[46] Gostein M, Stueve B, Chan M. Accurately measuring PV soiling losses with soiling station employing PV module power measurements. 2017 IEEE 44th Photovolt. Spec. Conf. PVSC 2017, 2017. doi:10.1109/PVSC.2017.8366169.

[47] Smith SW. The scientist and engineer's guide to digital signal processing. California Technical Pub; 1997.

[48] Gil R, Monné C, Bernal N, Muñoz M, Moreno F. Thermal model of a dish stirling cavity-receiver. Energies 2015;8:1042–57. doi:10.3390/en8021042.

[49] Eldighidy SM. Optimum outlet temperature of solar collector for maximum work output for an Otto air-standard cycle with ideal regeneration. Sol Energy 1993;51:175–82. doi:10.1016/0038-092X(93)90094-5.

[50] Hogan RE. AEETES-A solar reflux receiver thermal performance numerical model. Sol Energy 1994. doi:10.1016/0038-092X(94)90066-3.

[51] Yaqi L, Yaling H, Weiwei W. Optimization of solar-powered Stirling heat engine with finite-time thermodynamics. Renew Energy 2011;36:421–7. doi:10.1016/j.renene.2010.06.037.

[52] García D, González MA, Prieto JI, Herrero S, López S, Mesonero I, et al. Characterization of the power and efficiency of Stirling engine subsystems. Appl Energy 2014;121:51–63. doi:10.1016/j.apenergy.2014.01.067.

[53] Kallos G, Papadopoulos A, Katsafados P, Nickovic S. Transatlantic Saharan dust transport: Model simulation and results. J Geophys Res 2006;111:D09204. doi:10.1029/2005JD006207.

[54] Kallos G, Katsafados P, Spyrou C, Papadopoulos A. Desert dust deposition over the Mediterranean Sea estimated with the SKIRON/Eta - System validation. 4th EuroGOOS Conf., 2005.

[55] Querol X, Pey J, Pandolfi M, Alastuey A, Cusack M, Pérez N, et al. African dust contributions to mean ambient PM10 mass-levels across the Mediterranean Basin. Atmos Environ 2009;43:4266–77.

[56] Kallos G, Spyrou C, Astitha M, Mitsakou C, Solomos S, Kushta J, et al. Ten-year operational dust forecasting--Recent model development and future plans. IOP Conf. Ser. Earth Environ. Sci., vol. 7, 2009, p. 12012.

[57] Mitsakou C, Kallos G, Papantoniou N, Spyrou C, Solomos S, Astitha M, et al. Saharan dust levels in Greece and received inhalation doses. Atmos Chem Phys 2008;8:7181–92. doi:10.5194/acp-8-7181-2008.

[58] Astitha M, Kallos G. Gas-phase and aerosol chemistry interactions in South Europe and the Mediterranean region. Environ Fluid Mech 2009;9:3–22.

[59] National & Kapodistrian University of Athens School of Physics. SKIRON/Dust system n.d. http://forecast.uoa.gr/dustindx.php?domain=med (accessed October 10, 2019).

[60] Solomos S, Kallos G, Kushta J, Astitha M, Tremback C, Nenes A, et al. An integrated modeling study on the effects of mineral dust and sea salt particles on clouds and precipitation. Atmos Chem Phys 2011;11:873–92. doi:10.5194/acp-11-873-2011.